\documentclass[reprint,superscriptaddress,amsmath,amssymb,aps,prx,floatfix]{revtex4-2}

\usepackage[utf8]{inputenc}
\usepackage[T1]{fontenc}

\usepackage{graphicx}
\usepackage[dvipsnames]{xcolor}

\usepackage{hyperref}
\hypersetup{
  colorlinks,
  allcolors=NavyBlue,
  linktoc=all
}

\usepackage{soul}

\newcommand*{\unit}[1]{\ensuremath{\,\mathrm{#1}}}
\newcommand*{\s}[1]{\ensuremath{_\mathrm{#1}}}	
\newcommand*{\FSR}{\ensuremath{\nu_\mathrm{FSR}}}

\let\Re\relax
\let\Im\relax
\DeclareMathOperator{\Re}{Re}
\DeclareMathOperator{\Im}{Im}

\DeclareUnicodeCharacter{03B1}{\alpha}
\DeclareUnicodeCharacter{03B2}{\beta}
\DeclareUnicodeCharacter{03B3}{\gamma}
\DeclareUnicodeCharacter{03B4}{\delta}
\DeclareUnicodeCharacter{03B5}{\epsilon}
\DeclareUnicodeCharacter{03B6}{\zeta}
\DeclareUnicodeCharacter{03B7}{\eta}
\DeclareUnicodeCharacter{03B8}{\theta}
\DeclareUnicodeCharacter{03B9}{\iota}
\DeclareUnicodeCharacter{03BA}{\kappa}
\DeclareUnicodeCharacter{03BB}{\lambda}
\DeclareUnicodeCharacter{03BC}{\mu}
\DeclareUnicodeCharacter{03BD}{\nu}
\DeclareUnicodeCharacter{03BE}{\xi}
\DeclareUnicodeCharacter{03C0}{\pi}
\DeclareUnicodeCharacter{03C1}{\rho}
\DeclareUnicodeCharacter{03C3}{\sigma}
\DeclareUnicodeCharacter{03C4}{\tau}
\DeclareUnicodeCharacter{03C6}{\phi}
\DeclareUnicodeCharacter{03C7}{\chi}
\DeclareUnicodeCharacter{03C8}{\psi}
\DeclareUnicodeCharacter{03C9}{\omega}

\DeclareUnicodeCharacter{0393}{\Gamma}
\DeclareUnicodeCharacter{0394}{\Delta}
\DeclareUnicodeCharacter{03A6}{\Phi}
\DeclareUnicodeCharacter{03A9}{\Omega}

\DeclareUnicodeCharacter{210F}{\hbar}
\DeclareUnicodeCharacter{2020}{\dagger}
\DeclareUnicodeCharacter{2202}{\partial}

\begin{document}

\title{Route to hyperchaos in quadratic optomechanics}

\author{Lina Halef}
\author{Itay Shomroni}\email{itay.shomroni@mail.huji.ac.il}
\affiliation{ Racah Institute of Physics, The Hebrew University of Jerusalem, Jerusalem 9190401, Israel}

\date{April 17th, 2025}

\begin{abstract}
Hyperchaos is a qualitatively stronger form of chaos, in which several degrees of freedom contribute simultaneously to exponential divergence of small changes.
A hyperchaotic dynamical system is therefore even more unpredictable than a chaotic one, and has a higher fractal dimension.
While hyperchaos has been studied extensively over the last decades, only a few experimental systems are known to exhibit hyperchaotic dynamics.
Here we introduce hyperchaos in the context of cavity optomechanics, in which light inside an optical resonator interacts with a suspended oscillating mass.
We show that hyperchaos can arise in optomechanical systems with quadratic coupling
and is well within reach of current experiments.
We compute the two positive Lyapunov exponents, characteristic of hyperchaos, and independently verify the correlation dimension.
We also identify a possible mechanism for the emergence of hyperchaos.
As systems designed for high-precision measurements, optomechanical systems enable direct measurement of all four dynamical variables and therefore the full reconstruction of the hyperchaotic attractor.
Our results may contribute to better understanding of nonlinear systems and the chaos-hyperchaos transition, and allow the study of hyperchaos in the quantum regime.
\end{abstract}

\maketitle

Ultraprecise optical measurements of mechanical motion 
underlie pivotal technologies such as gravitational wave detectors~\cite{ligo2015} and
atomic~\cite{giessibl2003} and magnetic~\cite{rugar2004} force microscopes.
Light also carries momentum which can be used to manipulate mechanical objects, e.g.~in optical tweezers~\cite{ashkin1970}.
%
Cavity optomechanics~\cite{aspelmeyer2014} studies these interactions at the fundamental level,
employing a wide range of engineered mechanical oscillators,
in a setup where the oscillator is coupled to light enclosed in a high-quality optical cavity.
The main focus is the ensuing quantum dynamics, with the aim
to develop novel quantum technologies~\cite{barzanjeh2022}, search for dark matter~\cite{carney2021}, or address questions regarding quantum theory at the macroscopic scale~\cite{leggett2002}.

While most works focus on the linear regime of small oscillations,
the optomechanical interaction is fundamentally nonlinear
and can lead to rich dynamics.
The existence of multiple period-1 limit cycles
(sustained oscillations under a continuous optical drive)
was studied in~\cite{marquardt2006} and explored experimentally in~\cite{krause2015,buters2015}.
As the nonlinearity increases (e.g., by increasing the drive), a cascade of period-doubling bifurcations finally lead to chaos~\cite{bakemeier2015}.
Indeed, an early observation in optomechanics was that of chaotic mechanical motion~\cite{carmon2007}.
Sustained oscillations and chaos in optomechanics were further analyzed in various regimes~\cite{zhang2014,zhang2017,wurl2016,mumford2015,lu2015,roque2020,zhang2020,christou2021} including in the quantum regime~\cite{ludwig2008,schulz2016,bakemeier2015}.
Experimental work is scant, however, and often relies on other nonlinearities such as thermal effects or Duffing nonlinearity~\cite{monifi2016,navarro-urrios2017,wu2017,madiot2021}.

In this work we introduce a novel and rare feature in the dynamics of optomechanical systems: hyperchaos.
A~hallmark of chaos is sensitivity to initial conditions. In most dynamical systems, however, this sensitivity is dominated by a single dimension in phase space.
In other words, an infinitesimal volume on a chaotic attractor will tend to stretch at an exponential rate
along a single dimension transverse to the orbit.
While ubiquitous, this form of chaos is the lowest stage in a hierarchy of chaotic dynamics~\cite{rossler1983}---since it is possible that the expansion occur along multiple dimensions.
Such `hyperchaotic' behavior~\cite{rossler1979}
has been studied theoretically over the last decades~\cite{kapitaniak1991,harrison1999,kapitaniak2000,pavlov2015,du2018,leo_kingston2023}.
Despite this interest, only a handful of experiments demonstrated hyperchaotic dynamics, among which are electronic circuits~\cite{matsumoto1986}, an NMR system~\cite{stoop1988}, a semiconductor~\cite{stoop1989}, a certain chemical reaction~\cite{eiswirth1992}, lasers with delayed feedback~\cite{fischer1994,deng2022}, and optical solitons~\cite{bir2020}.
Unfortunately, with the exception of electronic circuits, in all these experiments
both the governing equations and the dimension of phase space are unknown.
Moreover, most dynamical variables
cannot be measured in the experiment,
and the structure of phase space must be inferred indirectly using
techniques which are susceptible to noise, and greatly exaggerate its dimensionality~\cite{abarbanel1993}.
This severely limits the insight that can be gained by studying these systems.
Hyperchaos has also been recently predicted in the polarization of vertical-cavity surface-emitting lasers~\cite{bonatto2018}, semiconductor superlattices~\cite{mompo2021},
hydromagnetic convection~\cite{macek2014}
and coupled two-level systems~\cite{andreev2021}%
---all of which also suffer from lack of full experimental access to the dynamical variables.

In contrast, in optomechanical systems all degrees of freedom can be measured with high precision.
In addition, the phase space has only four dimensions, the minimum required for hyperchaos, making it considerably simpler than the preceding examples.
Another attractive advantage is the ability to explore hyperchaos in the quantum regime.
We identify regimes where hyperchaotic dynamics occur in optomechanical systems \emph{with quadratic coupling}~\cite{sankey2010,paraiso2015,kaviani2015,burgwal2020,bullier2021,burgwal2023}.
%
%
%
%
Hyperchaos arises when the dissipation is weak, with the cavity decay rate lower than the mechanical frequency, also known as the resolved-sideband regime.

\begin{figure}
	\includegraphics[scale=1]{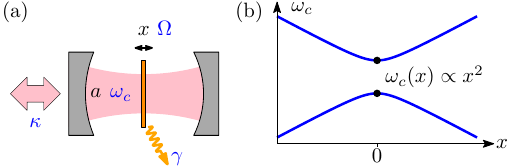}
	\caption{Quadratic optomechanical coupling.
		(a)~Membrane-in-the-middle system.
		A thin dielectric membrane with natural frequency $Ω$ oscillates with displacement $x$ within the mode $a$ of a Fabry-Perot cavity, shifting the cavity resonance frequency $ω_c$.
        The cavity decay rate through the input/output port is $κ$ and the oscillator decay rate is $γ$.
		(b)~Dependence of $ω_c$ on $x$.
        Only part of the full mode spectrum is shown, with two widely separated branches, originating from the hybridization of the modes of the two half-cavities.
        Quadratic coupling occurs at $x=0$.}
	\label{fig:1}
\end{figure}

\paragraph*{The System.}
Optomechanical systems~\cite{aspelmeyer2014} are based on a parametric coupling between light inside an optical cavity and an integrated mechanical oscillator of mass $m$ and frequency $Ω$.
The oscillator displacement $x$ shifts the resonance frequency $ω_c$ of the cavity by changing the optical path of the light, which in turn exerts a force on the oscillator through radiation pressure.
In quadratic optomechanical systems, some form of symmetry nullifies the first-order dependence of $ω_c$ on $x$,
leading to a frequency shift proportional to $x^2$~\cite{sankey2010,paraiso2015,kaviani2015,burgwal2020,bullier2021,burgwal2023}.
The canonical example is the so-called membrane-in-the-middle (MIM) system~\cite{thompson2008,jayich2008,wilson2009}
shown in Fig.~\ref{fig:1}(a).
A thin (usually tens of nanometers) semitransparent membrane is placed inside a Fabry-Perot cavity.
If placed at the node or antinode of the modal intensity distribution, two normal modes appear, as shown in Fig.~\ref{fig:1}(b).
Typically, the low reflectance of the membrane ($\lesssim 0.5$) results in a large frequency separation between the modes $\gg Ω$, and the membrane motion does not induce scattering between them. Therefore, only the driven mode participates in the dynamics~\cite{cheung2011}.

The dynamics are then governed by the Hamiltonian
\begin{equation}
    H = \frac{p^2}{2m} + \frac{1}{2}mΩ^2x^2 + ℏ(ω_c + Gx^2)a^†a.
    \label{eq:hamiltonian}
\end{equation}
The first two terms describe the free evolution of the oscillator, with momentum $p$ conjugate to $x$.
The third term describes the cavity field mode, characterized by the photon creation and annihilation operators $a^†,a$ respectively.
The coupling constant $G=\frac{1}{2}∂^2 ω_c/∂x^2$ determines the strength of the interaction.
Note that positive or negative $G$ can be selected, depending on which of the two modes we choose to work with [Fig.~\ref{fig:1}(b)].

The Hamiltonian~\eqref{eq:hamiltonian} is formulated quantum-mechanically as is standard in cavity optomechanics.
It contains Planck's constant $ℏ$ due to our use of photons.
We assume that the cavity is pumped with a strong coherent laser drive of frequency $ω_l=ω_c+Δ$ and amplitude $E$.
The cavity and oscillator energy decay rates are $κ$ and $γ$, respectively.
By scaling time and $κ,γ,Δ$ by the mechanical frequency $Ω$
and defining the dimensionless quantities~\footnote{If $G<0$ the minus sign appears outside the square root in $\tilde x,\tilde p$. The equations of motion are invariant under $x\rightarrow -x$, $p\rightarrow -p$.}
\begin{equation}
    \tilde a = \frac{Ω}{2E}a, \quad
    \tilde x = \sqrt{\frac{G}{Ω}}x, \quad
    \tilde p = \sqrt{\frac{G}{m^2Ω^3}}p, \quad
    P = \frac{8ℏGE^2}{mΩ^4}
\end{equation}
we obtain the classical equations of motion in a frame rotating with the drive frequency $ω_l$ (removing the tildes from $x,p,a$ for clarity)
\begin{subequations}
    \label{eq:eom}
    \begin{align}
        \dot x &= p \\
        \label{eq:dotp}
        \dot p &= -(1 + P\lvert a\rvert^2)x - γp\\
        \dot a &= i(Δ - x^2)a -\frac{κ}{2}a + \frac{1}{2}.
    \end{align}
\end{subequations}
Equations~\eqref{eq:eom} are nonlinear and cannot be solved analytically.
There are four parameters: the cavity and oscillator decay rates $κ,γ$ are typically fixed by the experimental system, while the pump $P$, which sets the strength of the nonlinearity, and the detuning $Δ$ can be tuned easily by adjusting the pumping laser.
The oscillator typically has a very high quality factor, $γ\ll 1$.

It is well known that both linear and quadratic coupling optomechanical systems exhibit phenomena such as multistability~\cite{marquardt2006}, period-doubling bifurcations, and chaos~\cite{bakemeier2015,roque2020,wurl2016}.
We note that since $a$ is complex, there are four equations in real variables, which is the minimum required for hyperchaos.

Equation~\eqref{eq:dotp} shows that the quadratic coupling shifts the squared mechanical frequency in proportion to the light intensity $\lvert a\rvert^2$.
If $P<0$, the frequency of the oscillator may become imaginary, 
leading to diverging solutions. 
Here we focus exclusively on $P>0$, corresponding to the upper branch of Fig.~\ref{fig:1}(b) where $ω_c$ has a minimum at $x=0$.
In this case the system has a single fixed point (where all time derivatives are zero), $x=p=0, a=1/(κ-2iΔ)$.
This fixed point is stable for all parameter values, hence transitions to self-sustained oscillations (limit cycles),
and later to chaos,
cannot occur through Hopf bifurcations
as in other optomechanical systems~\cite{bakemeier2015,wurl2016},
and need to be excited externally, i.e.~by setting the initial condition of the oscillator.
In an experiment, the oscillator can be initialized in the desired state by first applying a pump tuned to the lower branch of Fig.~\ref{fig:1}(b),
where $P<0$ and other fixed points exist, or by mechanical excitation.

Equations~\eqref{eq:eom}, with the addition of linear $x$ coupling, have been analyzed in previous works.
When $κ\gg 1$, the cavity field can be adiabatically eliminated, resulting in an effective potential for the oscillator~\cite{buchmann2012,seok2013,seok2013}.
Self-sustained oscillations and ordinary chaotic motion were studied in~\cite{zhang2014,zhang2017} for $κ\sim 1$ and small detunings $\lvert Δ\rvert\sim 1$.

\paragraph*{Chaos and Hyperchaos.}
A trajectory in phase space of a dynamical system is characterized by its spectrum of  Lyapunov exponents (LEs), the number of which equals the dimension of the phase space~\cite{strogatz2015}.
The LEs quantify the divergence of infinitesimally nearby trajectories, hence the sensitivity to initial conditions.
A single positive LE means that nearby trajectories separate exponentially fast, and is sufficient to classify the motion as chaotic.
If more than one LE is positive, however, then exponential divergence can occur along mutually independent directions, which indicates higher complexity of motion and is termed `hyperchaos'~\cite{rossler1979}.


\begin{figure}
    \includegraphics[scale=1]{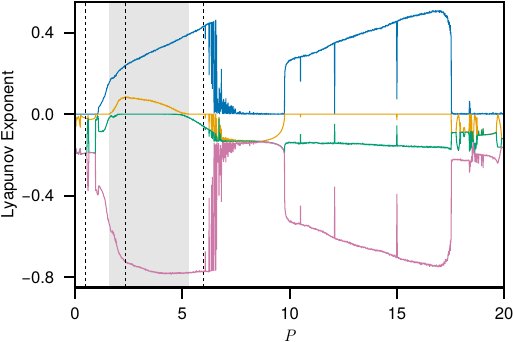}
    \caption{%
        The Lyapunov exponents of a system with $κ=0.4$, $γ=10^{-3}$, and $Δ=15.5$, for varying $P$.
        The initial condition is $(x,p,a)=(4,4,0)$.
        The shaded region highlights the range with two positive LEs, indicating hyperchaos.
    }
    \label{fig:lyapunov}
\end{figure}

\begin{figure}
    \includegraphics[scale=1]{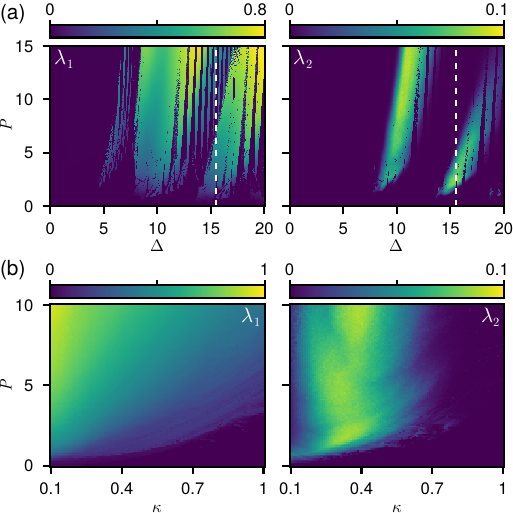}
    \caption{%
        Occurrence of hyperchaos for various system parameters.
        (a)~Values of the maximal LE $λ_1$ and second LE $λ_2$ vs.~the detuning $Δ$ and pump power $P$, with fixed cavity decay rate $κ=0.4$.
        The dashed white lines correspond to Fig.~\ref{fig:lyapunov}(a).
        (b)~Effect of the linewidth $κ$ for different pumping powers $P$,
        where the colormaps show, each independently, the largest values of $λ_1, λ_2$ obtained in the range $Δ \in [0,20]$ for a given $(κ,P)$.
        The other parameters are $γ=10^{-3}$ and initial condition $(x,p,a)=(4,4,0)$.}
    \label{fig:colormap}
\end{figure}

Figure~\ref{fig:lyapunov} shows the four LEs of the system~\eqref{eq:eom} with
$κ=0.4$, $γ=10^{-3}$, $Δ=15.5$, and varying $P$.
The initial condition is $(x,p,a)=(4,4,0)$.
In the region $1.6 < P < 5$ two LEs are positive, indicating hyperchaotic behavior in this parameter range.
Furthermore
for $9.8<P<17.5$ only one LE is positive, i.e., the motion is merely chaotic.
At $P\simeq 2.36$ the second largest LE $λ_2$ attains its maximal value, and
the two positive LEs are $(λ_1,λ_2)\simeq (0.240,0.085)$.
We also investigate the occurrence of hyperchaos for different parameters.
Figure~\ref{fig:colormap}(a) fixes $κ=0.4$ and shows the two largest LEs $λ_{1,2}$ as a function of $Δ$ and $P$.
The effect of the cavity linewidth $κ$ is shown in Fig.~\ref{fig:colormap}(b).
Here, we check for hyperchaos by taking the largest value attained by $λ_{1,2}$ for any $Δ \in [0,20]$.
Figure~\ref{fig:colormap}(b) shows that hyperchaos occurs for $κ\lesssim 0.6$, i.e.~in the resolved-sideband regime of optomechanics.
%
%
%
A recent work~\cite{roque2020} characterized this regime as weakly dissipative, and showed
that it promotes chaotic behavior in optomechanics with linear $x$ coupling.
Here we extend this to hyperchaos under $x^2$ coupling.

It is noteworthy that optical cavities are rarely used with such large detunings, $Δ/κ\approx 40$.
In the absence of the oscillator, at this detuning only $\sim 1.5\times 10^{-4}$ of the maximal energy (for resonant pumping) would be stored in the cavity.
Nevertheless, for sufficiently high power (and appropriate initial conditions) both the oscillator and field coordinates approach their on-resonance values. See Appendix~\ref{sec:trajectories} for sample trajectories.

\paragraph*{Fractal Dimension.}
We now turn to the important question of how to distinguish hyperchaos from chaos in an experiment.
While chaotic motion is easy to distinguish from a non-chaotic one, for example by observing the power spectral density of the cavity output light~\cite{carmon2007},
verifying hyperchaotic motion is not as straightforward.
One conventional approach is to estimate the LEs from an experimentally acquired time series~\cite{wolf1985,bryant1990}.
In most experiments, only a subset of phase space can be measured directly, but the LEs (and other invariants) can be evaluated by embedding the data in higher-dimensional spaces using time delays.
However, this approach may not be optimal for LEs other than the maximal due to measurement noise~\cite{abarbanel1993}.

In contrast, a remarkable advantage of optomechanical systems is experimental access to all four dynamical variables.
The quadratures of the intracavity field $a$ can be measured directly via heterodyne detection of the cavity output, and the position $x$ and momentum $p$ of the oscillator can be measured using a weak probe laser of a different wavelength, where linear coupling is dominant.
Hence the LEs can be calculated more accurately, with no need for time-delay embedding.

\begin{figure}
    \includegraphics[scale=1]{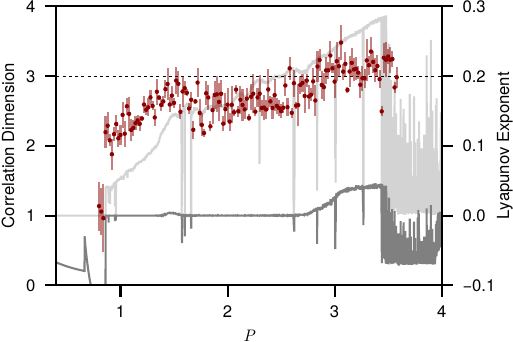}
    \caption{Estimation of the fractal dimension, using the corrlation dimension.
        System parameters are $κ=0.4$, $γ=0.007$, and $Δ=15.5$.
        The data (red, left axis) is overlaid on the first two Lyapunov exponents (right axis).
        There is excellent correlation between the second LE (dark gray) becoming positive, indicating hyperchaos, and the dimension exceeding 3, indicated by the dashed line.
        The dimension is obtained using the Grassberger-Procaccia method.
        Error bars indicate the standard error of a linear fit to the scaling curve, see Appendix~\ref{sec:numerics}.
        Note the increase of $D$ in the small hyperchaotic region near $P=1.5$.}
    \label{fig:corrdim}
\end{figure}

An attractor is also classified by its fractal dimension $D$, measuring its degree of self-similarity.
For an ordinary chaotic attractor $2<D<3$, while for hyperchaos $D>3$.
%
We use the standard method to estimate $D$ of simulated trajectories using the correlation dimension method of Grassberger and Proccacia~\cite{grassberger1983}.
This estimate is independent of the underlying equations and can therefore corroborate the existence of hyperchaos.
Figure~\ref{fig:corrdim} shows the estimated fractal dimension $D$
of a hyperchaotic system with $κ=0.4$, $γ=0.007$, $Δ=15.5$ and varying $P$ (see Appendix~\ref{sec:numerics} for numerical details).
We have chosen a larger mechanical decay rate in order to reduce the transient time of the dynamics, $\sim γ^{-1}$.
There is an excellent agreement between the hyperchaotic region with two positive LEs and $D>3$, and the chaotic region with $D\lesssim 3$.
To the best of our knowledge, estimation of $D$ for an experimentally-relevant hyperchaotic system was not carried out before.
%
In fact,
obtaining the fractal dimension posed severe computational requirements,
which gives a key clue as to the route to hyperchaos in our system, which we now discuss.

\paragraph*{Mechanism for Hyperchaos.}
It is well-known that the calculation of chaotic attractors is hindered by round-off errors, introduced by finite machine precision, which by definition diverge exponentially~\cite{sauer1997}.
Often this is disregarded,
and the computed trajectories considered true,
due to \emph{the shadowing lemma}~\cite{grebogi1990}, which states that the computed trajectory remains close to a different true trajectory, with different initial conditions (in other words, remains on the attractor).
But in general, many chaotic systems do not satisfy the prerequisites of the shadowing lemma~\cite{grebogi1990}.
In particular, hyperchaotic systems are believed to be unshadowable~\cite{mccullen2011}.
This means that to compute the trajectory to arbitrary time, one must use arbitrary precision.

We have found that this holds for our system.
Computing the correlation dimension based on different final times, with different solvers and/or different tolerances, yields very different values, which hence were not a good estimate of the true fractal dimension of the attractor.
To tackle this, we solved for the hyperchaotic trajectory with the largest $λ_2$
using arbitrary precision arithmetic with increasingly stringent tolerances, until the computed correlation dimension scaling curves were numerically identical.
This yielded absolute (relative) tolerance of $10^{-70}$ ($10^{-67}$).
Such extreme tolerances required using Feagin's 14th order explicit Runge-Kutta solver.

While the chaos-hyperchaos transition is still not well-understood in general,
unshadowability in our system provides strong evidence for a specific mechanism~\cite{davidchack2000}.
It is well-known that a chaotic attractor embeds an infinite number of \emph{unstable periodic orbits} (UPOs) that constitute its `skeleton'~\cite{auerbach1987,dawson1996}.
As a system parameter changes, increasingly more UPOs that are initially unstable in one dimension, bifurcate and become unstable along additional dimensions.
Thus the supported chaotic attractor transitions from chaos to hyperchaos~\cite{davidchack2000}.
Further evidence that this mechanism may be in place is the smooth change of $λ_2$ to positive values (Fig.~\ref{fig:lyapunov}).
Such transition is also accompanied by a phenomenon known as \emph{unstable dimension variability} (UDV)~\cite{dawson1994,dawson1996,davidchack2000}: within the transition region, the local dimension fluctuates as the system traverses parts of the attractor near UPOs with varying degrees of instablity.
UDV has the consequence of making the system nonhyperbolic and thus unshadowable~\cite{davidchack2000,lai1999,barreto2000}.

The above mechanism has been studied in systems of two coupled nonlinear oscillators, each capable of chaotic behavior (i.e., six dimensions in total), and attributed to degrees of chaos syncronization between the oscillators~\cite{lai1999,mccullen2011,kuptsov2013}.
To the best of our knowledge, it has not been shown in four-dimensional systems, which are not decomposable in this manner.

\begin{figure*}
    \includegraphics[scale=1]{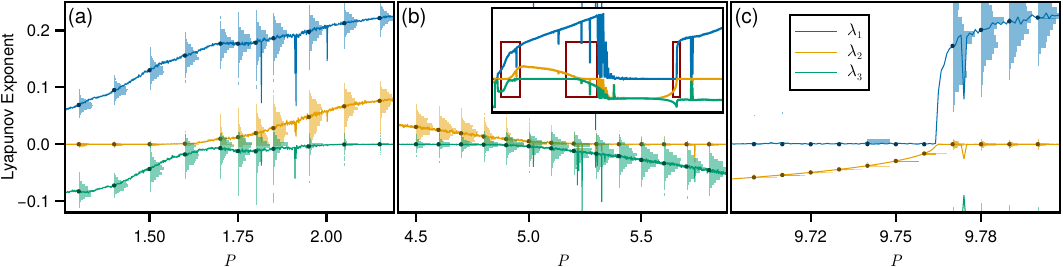}
    \caption{%
        Fluctuations of the LEs, highlighting unstable dimension variability at the chaos-hyperchaos transition.
        Parameters are the same as in Fig.~\ref{fig:lyapunov}.
        The three largest LEs are computed for finite (but large) time intervals and their distributions, at points denoted by markers, is shown on top of the asymptotic LEs.
        The distributions are half-transparent, revealing the overlap of $λ_2$ and $λ_3$;
        their vertical axis corresponds to the LE value and the horizontal is arbitrary.
        (a) and (b) show chaos-hyperchaos transition, while (c) shows transition from non-chaotic behavior to regular chaos.
        Note that at some points,
        particularly in~(c),
        the distributions of $λ_2$ are narrow and hardly visible.
        The inset in~(b) shows a wider view, similar to Fig.~\ref{fig:lyapunov}, indicating  the regions in~(a), (b), and~(c), left to right respectively.
        }
    \label{fig:FTLE}
\end{figure*}

One may verify UDV and further justify this mechanism by examining the finite-time LEs (FTLEs)~\cite{dawson1994,dawson1996}.
We evolve the system for $10^7$ time steps subdivided into $10^3$ intervals, and compute the LEs for each interval.
Figure~\ref{fig:FTLE}(a,b) shows the distributions of these FTLEs along the chaos-hyperchaos transition.
As the transition is approached, the negative $λ_3$, denoting directions along which the trajectory is still stable, starts to fluctuate wildly.
The fluctuations pass to $λ_2$ (initially close to zero---along the trajectory), at which point the two LEs coalesce, fluctuating about zero and hard to distinguish.
Note that such behavior is \emph{not} encountered in transition to regular chaos [Fig.~\ref{fig:FTLE}(c)]: at no point does the LE fluctuate about zero.

\paragraph*{Experimental Relevance.}
The main challenge in observing hyperchaos in an experiment is the optical power needed to achieve the required value of $P$.
Too high pumping power will lead to deleterious effects such as excess heating that will hinder the measurement.
In addition, the cavity mirrors should have the same reflectivity, to avoid dissipative coupling~\cite{elste2009,yanay2016}.

For a MIM system, the optomechanical coupling rate is evaluated from the cavity response to the position $x$, given by
$ω_c(x) = 2\FSR\arccos[r_d \cos(2kx)]$,
where $k=2π/λ$ is the wavenumber for light of wavelength $λ$, $\FSR=c/2L$ is the free spectral range of a cavity with length $L$, and $r_d$ the amplitude reflectivity of the membrane~\cite{jayich2008}.
Pure quadratic coupling occurs when $dω_c/dx = 0$, where
\begin{equation}
    G = \frac{1}{2}\frac{d^2ω_c}{dx^2} = 4k^2\FSR\frac{r_d}{\sqrt{1-r_d^2}}.
\end{equation}


The pumping amplitude $E$ can be written as $\sqrt{κ\s{ex}}a\s{in}$ where $\lvert a\s{in}\rvert^2$ is the input photon flux and $κ\s{ex}$ is the cavity input coupling rate.
We assume $κ\s{ex}=κ$ for simplicity.
Collecting all the parameters, we can express the required incident power on the cavity for a given $P$ as
\begin{equation}
    \frac{\text{incident power}}{P} = \frac{Ω^4 λmL}{32πκ\s{ex}}\sqrt{\frac{1-r_d^2}{r_d^2}}
\end{equation}

For example, Ref.~\onlinecite{reinhardt2016} reports a membrane with $Ω=2π\times 101.9\unit{kHz}$, $m=3\times 10^{-12}\unit{kg}$, and $r_d = 0.58$ at $λ=532\unit{nm}$.
Setting $L=30\unit{mm}$ and $κ = 0.4Ω$ yields a `modest' cavity finesse of $20,000$ and only $1\unit{mW}$ of pump power to achieve $P=2.36$ required for observation of hyperchaos.
For these parameters, $x=1$ corresponds to 8 times the r.m.s of thermal fluctuations of the oscillator at room temperature.
Optomechanical systems using levitated nanospheres~\cite{fonseca2016,delic2020} are also attractive, due to the small mass $m\sim 10^{-17}\unit{kg}$ and mechanical frequencies $Ω/2π\sim 10\text{--}100\unit{kHz}$ set by an external potential.
We conclude that observation of hyperchaotic optomechanical motion is very feasible with current systems.

\paragraph*{Conclusion.}
We have shown that hyperchaos exists in quadratic optomechanics and is observable with current experimental systems.
It is important to note that while we focused here on pure quadratic coupling, our calculations show that hyperchaos exists also in the presence of additional linear coupling.
Furthermore, while we presented a system with the minimal required dimensionality for hyperchaos, this can be extended by incorporating additional optical and mechanical modes.
Another interesting prospect is hyperchaos in the quantum regime, which is routinely accessible in optomechanical systems.

We also outlined a possible mechanism for the emergence of hyperchaos.
The chaos-hyperchaos transition remains an active field of research~\cite{pavlov2015,du2018,leo_kingston2022} with implications to a wide range of scientific fields~\cite{garashchuk2019,sajjadi2020,lin2020}.
Optomechanics can serve as an ideal experimental platform for the study of these phenomena.

\paragraph*{Data Availability.}
The code and data used to produce the plots within this paper are available
from the corresponding author upon reasonable request.

\begin{acknowledgments}
We thank Omri Gat, Snir Gazit, and Nils J.~Engelsen for fruitful discussions.
This work was supported by the Israel Science Foundation Grant No.~695/22.
L.H. acknowledges support from the Kaete Klausner Scholarship.
\end{acknowledgments}

\clearpage

\appendix

\section{Equations of motion}
\label{sec:eom}

For computation, Eqs.~\eqref{eq:eom} can be recast using real variables.
Setting $a_x = \Re{a}$ and $a_p = \Im{a}$ we have
\begin{subequations}
    \label{eq:eomreal}
    \begin{align}
        \dot x &= p \\
        \dot p &= -x - P(a_x^2 + a_p^2) x - γp\\
        \dot a_x &= -Δ a_p + x^2 a_p -\frac{κ}{2}a_x + \frac{1}{2} \\
        \dot a_p &= Δ a_x - x^2 a_x -\frac{κ}{2}a_p
    \end{align}
\end{subequations}
The fixed points of the system are found by setting the time derivatives in~\eqref{eq:eom} to zero.
This yields that
if $P>0$, the only fixed point is $x=p=0$, $a=(κ-2iΔ)^{-1}$.
It can be shown that at this fixed point, the eigenvalues of the Jacobian of the system~\eqref{eq:eomreal} always have a negative real part
(for $P>0$ and~$γ<2$) making this fixed point stable.

For $P<0$ there are additional fixed points with $x\neq 0$ and $\lvert a\rvert^2=-1/P$.
These fixed points exist if 
$μ \equiv \sqrt{-P-κ^2}$ is positive, then 
\begin{equation}
    x = \pm\sqrt{Δ\pm\frac{μ}{2}}
\end{equation}
which exist when the radicand is positive.
For all fixed points with $x\neq 0$, we have
\begin{equation}
    a = \frac{1}{κ-iμ} = \frac{1}{\sqrt{-P}}e^{iφ}, \quad \text{where }\cos φ = \frac{κ}{\sqrt{-P}}.
\end{equation}

\section{Numerical Details}
\label{sec:numerics}

\begin{figure}[t]
    \includegraphics[scale=1]{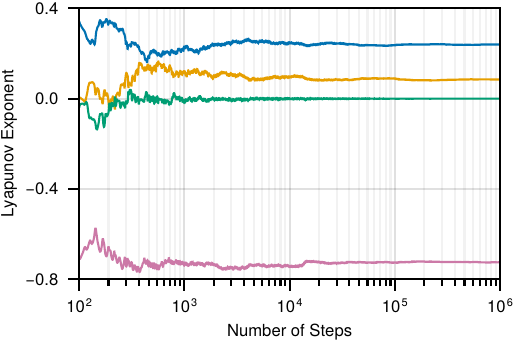}
    \caption{%
        Convergence of the Lyapunov exponents.
        Calculated with parameters $\{P,Δ,κ,γ\} = \{2.36,15.5,0.4,0.001\}$ and initial condition $(x,p,a)=(4,4,0)$.
        The time step between orthogonalizations is $Δt=0.1$.}
    \label{fig:convergence}
\end{figure}

\paragraph*{Convergence of the Lyapunov exponents.}
Numerical simulations were done in Julia using the ChaosTools package~\cite{datseris2022}.
The values of the LEs were verified in Python using the lyapynov package
and by hand-coding Benettin's algorithm~\cite{benettin1980,benettin1980b}.
Figure~\ref{fig:convergence} shows the convergence of the computed LEs as a function of the number of steps.

\paragraph*{Correlation dimension.}
For the computation of the correlation dimension in Fig.~\ref{fig:corrdim},
we integrated the dynamical equations for each $P$ using the DifferentialEquations Julia package with arbitrary precision arithmetic (Julia's BigFloat).
The system was propagated from $t=0$ to $t=550$, discarding data up to $t=200$ to eliminate transients.
The trajectories were then standardized.
The correlation sum $C(ε)$ was calculated using the FractalDimensions Julia package~\cite{datseris2023}, the interpoint distance $ε$ ranging from one octave above to two octaves below its minimal/maximal values respectively.
A Theiler window of 70 time units was used, but other choices do not modify the results substantially.
The trajectories were simulated with a time step $dt=0.001$.
This gave a total of 350,000 points or $> 60\times 10^9$ pair distances for each trajectory.
To calculate $C(ε)$ for each trajectory, we divided it into ten interleaving trajectories with $dt=0.01$ and averaged $C(ε)$ over these.
This was done to further reduce temporal correlations but again does not impact the results.

The correlation dimension was estimated by a simple linear least-squares fit of $\log C$ vs.~$\log ε$, starting from values of $C(ε)$ resulting from at least 100,000 pairs, and its error by the standard error of this estimate.
Though it has been pointed out that such estimates are not completely accurate~\cite{judd1992}, we find it gives good agreement in our case.
A more involved method to compute the scaling region and correlation dimension was also employed~\cite{deshmukh2021}, and gave similar results.

\section{Sample Trajectories}
\label{sec:trajectories}

\begin{figure*}
    \includegraphics[scale=1]{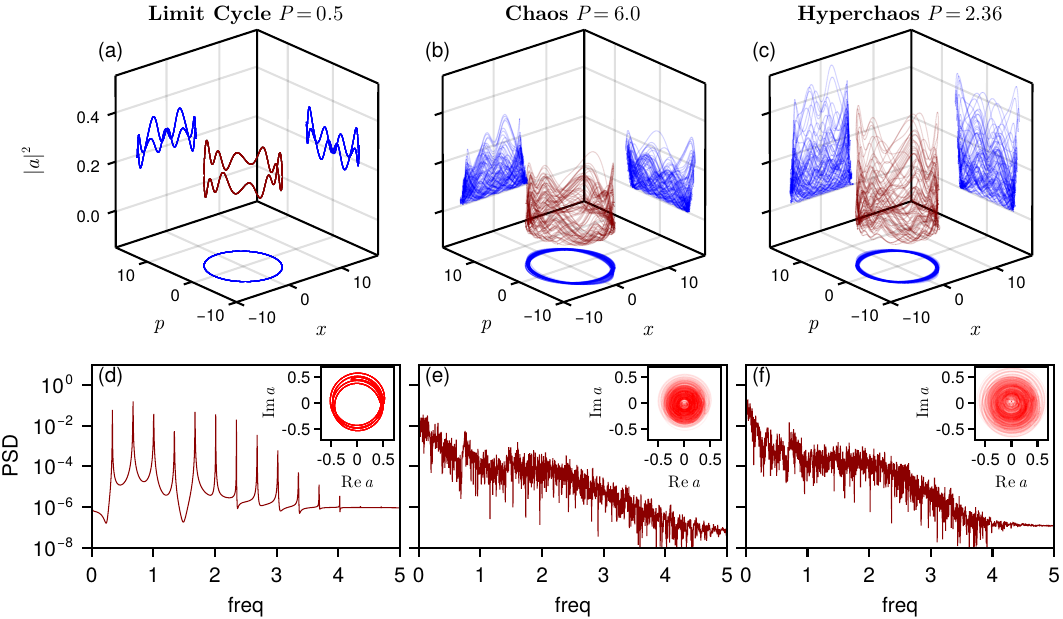}
    \caption{Phase space trajectories of a system with the same parameters as Fig.~\ref{fig:lyapunov}, for cases of a limit cycle (left column), chaos (middle column) and hyperchaos (right column).
        These parameters are also indicated with the dashed lines in Fig.~\ref{fig:lyapunov}.
        Panels (a--c) show the trajectories using the variables $x$, $p$, $\lvert a\rvert^2$, along with the projections of each variable pair.
        Panels (d--f) show the power spectral densities of the intracavity field $a$.
        The insets show the trajectories of the quadratures of $a$.}
    \label{fig:trajectories}
\end{figure*}

Figure~\ref{fig:trajectories} shows three selected trajectories, for the cases of a limit cycle, chaos, and hyperchaos, categorized by the Lyapunov spectrum and indicated by dashed lines in Fig.~\ref{fig:lyapunov}.
All the trajectories cover the same time interval and contain the same number of points.
In all cases, the dynamics of the light field $a$ are more complex than that of the oscillator [see also the insets of Fig.~\ref{fig:trajectories}(d--f)].

\clearpage

\bibliography{refs}

\begin{thebibliography}{89}%
\makeatletter
\providecommand \@ifxundefined [1]{%
 \@ifx{#1\undefined}
}%
\providecommand \@ifnum [1]{%
 \ifnum #1\expandafter \@firstoftwo
 \else \expandafter \@secondoftwo
 \fi
}%
\providecommand \@ifx [1]{%
 \ifx #1\expandafter \@firstoftwo
 \else \expandafter \@secondoftwo
 \fi
}%
\providecommand \natexlab [1]{#1}%
\providecommand \enquote  [1]{``#1''}%
\providecommand \bibnamefont  [1]{#1}%
\providecommand \bibfnamefont [1]{#1}%
\providecommand \citenamefont [1]{#1}%
\providecommand \href@noop [0]{\@secondoftwo}%
\providecommand \href [0]{\begingroup \@sanitize@url \@href}%
\providecommand \@href[1]{\@@startlink{#1}\@@href}%
\providecommand \@@href[1]{\endgroup#1\@@endlink}%
\providecommand \@sanitize@url [0]{\catcode `\\12\catcode `\$12\catcode
  `\&12\catcode `\#12\catcode `\^12\catcode `\_12\catcode `\%12\relax}%
\providecommand \@@startlink[1]{}%
\providecommand \@@endlink[0]{}%
\providecommand \url  [0]{\begingroup\@sanitize@url \@url }%
\providecommand \@url [1]{\endgroup\@href {#1}{\urlprefix }}%
\providecommand \urlprefix  [0]{URL }%
\providecommand \Eprint [0]{\href }%
\providecommand \doibase [0]{https://doi.org/}%
\providecommand \selectlanguage [0]{\@gobble}%
\providecommand \bibinfo  [0]{\@secondoftwo}%
\providecommand \bibfield  [0]{\@secondoftwo}%
\providecommand \translation [1]{[#1]}%
\providecommand \BibitemOpen [0]{}%
\providecommand \bibitemStop [0]{}%
\providecommand \bibitemNoStop [0]{.\EOS\space}%
\providecommand \EOS [0]{\spacefactor3000\relax}%
\providecommand \BibitemShut  [1]{\csname bibitem#1\endcsname}%
\let\auto@bib@innerbib\@empty
\bibitem [{\citenamefont {{The LIGO Scientific Collaboration \textit{et
  al.}}}(2015)}]{ligo2015}%
  \BibitemOpen
  \bibfield  {author} {\bibinfo {author} {\bibnamefont {{The LIGO Scientific
  Collaboration \textit{et al.}}}},\ }\bibfield  {title} {\bibinfo {title}
  {Advanced {LIGO}},\ }\href {https://doi.org/10.1088/0264-9381/32/7/074001}
  {\bibfield  {journal} {\bibinfo  {journal} {Classical and Quantum Gravity}\
  }\textbf {\bibinfo {volume} {32}},\ \bibinfo {pages} {074001} (\bibinfo
  {year} {2015})}\BibitemShut {NoStop}%
\bibitem [{\citenamefont {Giessibl}(2003)}]{giessibl2003}%
  \BibitemOpen
  \bibfield  {author} {\bibinfo {author} {\bibfnamefont {F.~J.}\ \bibnamefont
  {Giessibl}},\ }\bibfield  {title} {\bibinfo {title} {Advances in atomic force
  microscopy},\ }\href {https://doi.org/10.1103/RevModPhys.75.949} {\bibfield
  {journal} {\bibinfo  {journal} {Reviews of Modern Physics}\ }\textbf
  {\bibinfo {volume} {75}},\ \bibinfo {pages} {949} (\bibinfo {year}
  {2003})}\BibitemShut {NoStop}%
\bibitem [{\citenamefont {Rugar}\ \emph {et~al.}(2004)\citenamefont {Rugar},
  \citenamefont {Budakian}, \citenamefont {Mamin},\ and\ \citenamefont
  {Chui}}]{rugar2004}%
  \BibitemOpen
  \bibfield  {author} {\bibinfo {author} {\bibfnamefont {D.}~\bibnamefont
  {Rugar}}, \bibinfo {author} {\bibfnamefont {R.}~\bibnamefont {Budakian}},
  \bibinfo {author} {\bibfnamefont {H.~J.}\ \bibnamefont {Mamin}},\ and\
  \bibinfo {author} {\bibfnamefont {B.~W.}\ \bibnamefont {Chui}},\ }\bibfield
  {title} {\bibinfo {title} {Single spin detection by magnetic resonance force
  microscopy},\ }\href {https://doi.org/10.1038/nature02658} {\bibfield
  {journal} {\bibinfo  {journal} {Nature}\ }\textbf {\bibinfo {volume} {430}},\
  \bibinfo {pages} {329} (\bibinfo {year} {2004})}\BibitemShut {NoStop}%
\bibitem [{\citenamefont {Ashkin}(1970)}]{ashkin1970}%
  \BibitemOpen
  \bibfield  {author} {\bibinfo {author} {\bibfnamefont {A.}~\bibnamefont
  {Ashkin}},\ }\bibfield  {title} {\bibinfo {title} {Acceleration and
  {Trapping} of {Particles} by {Radiation} {Pressure}},\ }\href
  {https://doi.org/10.1103/PhysRevLett.24.156} {\bibfield  {journal} {\bibinfo
  {journal} {Physical Review Letters}\ }\textbf {\bibinfo {volume} {24}},\
  \bibinfo {pages} {156} (\bibinfo {year} {1970})}\BibitemShut {NoStop}%
\bibitem [{\citenamefont {Aspelmeyer}\ \emph {et~al.}(2014)\citenamefont
  {Aspelmeyer}, \citenamefont {Kippenberg},\ and\ \citenamefont
  {Marquardt}}]{aspelmeyer2014}%
  \BibitemOpen
  \bibfield  {author} {\bibinfo {author} {\bibfnamefont {M.}~\bibnamefont
  {Aspelmeyer}}, \bibinfo {author} {\bibfnamefont {T.~J.}\ \bibnamefont
  {Kippenberg}},\ and\ \bibinfo {author} {\bibfnamefont {F.}~\bibnamefont
  {Marquardt}},\ }\bibfield  {title} {\bibinfo {title} {Cavity optomechanics},\
  }\href {https://doi.org/10.1103/RevModPhys.86.1391} {\bibfield  {journal}
  {\bibinfo  {journal} {Reviews of Modern Physics}\ }\textbf {\bibinfo {volume}
  {86}},\ \bibinfo {pages} {1391} (\bibinfo {year} {2014})}\BibitemShut
  {NoStop}%
\bibitem [{\citenamefont {Barzanjeh}\ \emph {et~al.}(2022)\citenamefont
  {Barzanjeh}, \citenamefont {Xuereb}, \citenamefont {Gröblacher},
  \citenamefont {Paternostro}, \citenamefont {Regal},\ and\ \citenamefont
  {Weig}}]{barzanjeh2022}%
  \BibitemOpen
  \bibfield  {author} {\bibinfo {author} {\bibfnamefont {S.}~\bibnamefont
  {Barzanjeh}}, \bibinfo {author} {\bibfnamefont {A.}~\bibnamefont {Xuereb}},
  \bibinfo {author} {\bibfnamefont {S.}~\bibnamefont {Gröblacher}}, \bibinfo
  {author} {\bibfnamefont {M.}~\bibnamefont {Paternostro}}, \bibinfo {author}
  {\bibfnamefont {C.~A.}\ \bibnamefont {Regal}},\ and\ \bibinfo {author}
  {\bibfnamefont {E.~M.}\ \bibnamefont {Weig}},\ }\bibfield  {title} {\bibinfo
  {title} {Optomechanics for quantum technologies},\ }\href
  {https://doi.org/10.1038/s41567-021-01402-0} {\bibfield  {journal} {\bibinfo
  {journal} {Nature Physics}\ }\textbf {\bibinfo {volume} {18}},\ \bibinfo
  {pages} {15} (\bibinfo {year} {2022})}\BibitemShut {NoStop}%
\bibitem [{\citenamefont {Carney}\ \emph {et~al.}(2021)\citenamefont {Carney},
  \citenamefont {Krnjaic}, \citenamefont {Moore}, \citenamefont {Regal},
  \citenamefont {Afek}, \citenamefont {Bhave}, \citenamefont {Brubaker},
  \citenamefont {Corbitt}, \citenamefont {Cripe}, \citenamefont {Crisosto},
  \citenamefont {Geraci}, \citenamefont {Ghosh}, \citenamefont {Harris},
  \citenamefont {Hook}, \citenamefont {Kolb}, \citenamefont {Kunjummen},
  \citenamefont {Lang}, \citenamefont {Li}, \citenamefont {Lin}, \citenamefont
  {Liu}, \citenamefont {Lykken}, \citenamefont {Magrini}, \citenamefont
  {Manley}, \citenamefont {Matsumoto}, \citenamefont {Monte}, \citenamefont
  {Monteiro}, \citenamefont {Purdy}, \citenamefont {Riedel}, \citenamefont
  {Singh}, \citenamefont {Singh}, \citenamefont {Sinha}, \citenamefont
  {Taylor}, \citenamefont {Qin}, \citenamefont {Wilson},\ and\ \citenamefont
  {Zhao}}]{carney2021}%
  \BibitemOpen
  \bibfield  {author} {\bibinfo {author} {\bibfnamefont {D.}~\bibnamefont
  {Carney}}, \bibinfo {author} {\bibfnamefont {G.}~\bibnamefont {Krnjaic}},
  \bibinfo {author} {\bibfnamefont {D.~C.}\ \bibnamefont {Moore}}, \bibinfo
  {author} {\bibfnamefont {C.~A.}\ \bibnamefont {Regal}}, \bibinfo {author}
  {\bibfnamefont {G.}~\bibnamefont {Afek}}, \bibinfo {author} {\bibfnamefont
  {S.}~\bibnamefont {Bhave}}, \bibinfo {author} {\bibfnamefont
  {B.}~\bibnamefont {Brubaker}}, \bibinfo {author} {\bibfnamefont
  {T.}~\bibnamefont {Corbitt}}, \bibinfo {author} {\bibfnamefont
  {J.}~\bibnamefont {Cripe}}, \bibinfo {author} {\bibfnamefont
  {N.}~\bibnamefont {Crisosto}}, \bibinfo {author} {\bibfnamefont
  {A.}~\bibnamefont {Geraci}}, \bibinfo {author} {\bibfnamefont
  {S.}~\bibnamefont {Ghosh}}, \bibinfo {author} {\bibfnamefont {J.~G.~E.}\
  \bibnamefont {Harris}}, \bibinfo {author} {\bibfnamefont {A.}~\bibnamefont
  {Hook}}, \bibinfo {author} {\bibfnamefont {E.~W.}\ \bibnamefont {Kolb}},
  \bibinfo {author} {\bibfnamefont {J.}~\bibnamefont {Kunjummen}}, \bibinfo
  {author} {\bibfnamefont {R.~F.}\ \bibnamefont {Lang}}, \bibinfo {author}
  {\bibfnamefont {T.}~\bibnamefont {Li}}, \bibinfo {author} {\bibfnamefont
  {T.}~\bibnamefont {Lin}}, \bibinfo {author} {\bibfnamefont {Z.}~\bibnamefont
  {Liu}}, \bibinfo {author} {\bibfnamefont {J.}~\bibnamefont {Lykken}},
  \bibinfo {author} {\bibfnamefont {L.}~\bibnamefont {Magrini}}, \bibinfo
  {author} {\bibfnamefont {J.}~\bibnamefont {Manley}}, \bibinfo {author}
  {\bibfnamefont {N.}~\bibnamefont {Matsumoto}}, \bibinfo {author}
  {\bibfnamefont {A.}~\bibnamefont {Monte}}, \bibinfo {author} {\bibfnamefont
  {F.}~\bibnamefont {Monteiro}}, \bibinfo {author} {\bibfnamefont
  {T.}~\bibnamefont {Purdy}}, \bibinfo {author} {\bibfnamefont {C.~J.}\
  \bibnamefont {Riedel}}, \bibinfo {author} {\bibfnamefont {R.}~\bibnamefont
  {Singh}}, \bibinfo {author} {\bibfnamefont {S.}~\bibnamefont {Singh}},
  \bibinfo {author} {\bibfnamefont {K.}~\bibnamefont {Sinha}}, \bibinfo
  {author} {\bibfnamefont {J.~M.}\ \bibnamefont {Taylor}}, \bibinfo {author}
  {\bibfnamefont {J.}~\bibnamefont {Qin}}, \bibinfo {author} {\bibfnamefont
  {D.~J.}\ \bibnamefont {Wilson}},\ and\ \bibinfo {author} {\bibfnamefont
  {Y.}~\bibnamefont {Zhao}},\ }\bibfield  {title} {\bibinfo {title} {Mechanical
  quantum sensing in the search for dark matter},\ }\href
  {https://doi.org/10.1088/2058-9565/abcfcd} {\ \textbf {\bibinfo {volume}
  {6}},\ \bibinfo {pages} {024002} (\bibinfo {year} {2021})}\BibitemShut
  {NoStop}%
\bibitem [{\citenamefont {Leggett}(2002)}]{leggett2002}%
  \BibitemOpen
  \bibfield  {author} {\bibinfo {author} {\bibfnamefont {A.~J.}\ \bibnamefont
  {Leggett}},\ }\bibfield  {title} {\bibinfo {title} {Testing the limits of
  quantum mechanics: motivation, state of play, prospects},\ }\href
  {https://doi.org/10.1088/0953-8984/14/15/201} {\bibfield  {journal} {\bibinfo
   {journal} {Journal of Physics: Condensed Matter}\ }\textbf {\bibinfo
  {volume} {14}},\ \bibinfo {pages} {R415} (\bibinfo {year}
  {2002})}\BibitemShut {NoStop}%
\bibitem [{\citenamefont {Marquardt}\ \emph {et~al.}(2006)\citenamefont
  {Marquardt}, \citenamefont {Harris},\ and\ \citenamefont
  {Girvin}}]{marquardt2006}%
  \BibitemOpen
  \bibfield  {author} {\bibinfo {author} {\bibfnamefont {F.}~\bibnamefont
  {Marquardt}}, \bibinfo {author} {\bibfnamefont {J.~G.~E.}\ \bibnamefont
  {Harris}},\ and\ \bibinfo {author} {\bibfnamefont {S.~M.}\ \bibnamefont
  {Girvin}},\ }\bibfield  {title} {\bibinfo {title} {Dynamical {Multistability}
  {Induced} by {Radiation} {Pressure} in {High}-{Finesse} {Micromechanical}
  {Optical} {Cavities}},\ }\href
  {https://doi.org/10.1103/PhysRevLett.96.103901} {\bibfield  {journal}
  {\bibinfo  {journal} {Physical Review Letters}\ }\textbf {\bibinfo {volume}
  {96}},\ \bibinfo {pages} {103901} (\bibinfo {year} {2006})}\BibitemShut
  {NoStop}%
\bibitem [{\citenamefont {Krause}\ \emph {et~al.}(2015)\citenamefont {Krause},
  \citenamefont {Hill}, \citenamefont {Ludwig}, \citenamefont {Safavi-Naeini},
  \citenamefont {Chan}, \citenamefont {Marquardt},\ and\ \citenamefont
  {Painter}}]{krause2015}%
  \BibitemOpen
  \bibfield  {author} {\bibinfo {author} {\bibfnamefont {A.~G.}\ \bibnamefont
  {Krause}}, \bibinfo {author} {\bibfnamefont {J.~T.}\ \bibnamefont {Hill}},
  \bibinfo {author} {\bibfnamefont {M.}~\bibnamefont {Ludwig}}, \bibinfo
  {author} {\bibfnamefont {A.~H.}\ \bibnamefont {Safavi-Naeini}}, \bibinfo
  {author} {\bibfnamefont {J.}~\bibnamefont {Chan}}, \bibinfo {author}
  {\bibfnamefont {F.}~\bibnamefont {Marquardt}},\ and\ \bibinfo {author}
  {\bibfnamefont {O.}~\bibnamefont {Painter}},\ }\bibfield  {title} {\bibinfo
  {title} {Nonlinear {Radiation} {Pressure} {Dynamics} in an {Optomechanical}
  {Crystal}},\ }\href {https://doi.org/10.1103/PhysRevLett.115.233601}
  {\bibfield  {journal} {\bibinfo  {journal} {Physical Review Letters}\
  }\textbf {\bibinfo {volume} {115}},\ \bibinfo {pages} {233601} (\bibinfo
  {year} {2015})}\BibitemShut {NoStop}%
\bibitem [{\citenamefont {Buters}\ \emph {et~al.}(2015)\citenamefont {Buters},
  \citenamefont {Eerkens}, \citenamefont {Heeck}, \citenamefont {Weaver},
  \citenamefont {Pepper}, \citenamefont {de~Man},\ and\ \citenamefont
  {Bouwmeester}}]{buters2015}%
  \BibitemOpen
  \bibfield  {author} {\bibinfo {author} {\bibfnamefont {F.~M.}\ \bibnamefont
  {Buters}}, \bibinfo {author} {\bibfnamefont {H.~J.}\ \bibnamefont {Eerkens}},
  \bibinfo {author} {\bibfnamefont {K.}~\bibnamefont {Heeck}}, \bibinfo
  {author} {\bibfnamefont {M.~J.}\ \bibnamefont {Weaver}}, \bibinfo {author}
  {\bibfnamefont {B.}~\bibnamefont {Pepper}}, \bibinfo {author} {\bibfnamefont
  {S.}~\bibnamefont {de~Man}},\ and\ \bibinfo {author} {\bibfnamefont
  {D.}~\bibnamefont {Bouwmeester}},\ }\bibfield  {title} {\bibinfo {title}
  {Experimental exploration of the optomechanical attractor diagram and its
  dynamics},\ }\href {https://doi.org/10.1103/PhysRevA.92.013811} {\bibfield
  {journal} {\bibinfo  {journal} {Physical Review A}\ }\textbf {\bibinfo
  {volume} {92}},\ \bibinfo {pages} {013811} (\bibinfo {year}
  {2015})}\BibitemShut {NoStop}%
\bibitem [{\citenamefont {Bakemeier}\ \emph {et~al.}(2015)\citenamefont
  {Bakemeier}, \citenamefont {Alvermann},\ and\ \citenamefont
  {Fehske}}]{bakemeier2015}%
  \BibitemOpen
  \bibfield  {author} {\bibinfo {author} {\bibfnamefont {L.}~\bibnamefont
  {Bakemeier}}, \bibinfo {author} {\bibfnamefont {A.}~\bibnamefont
  {Alvermann}},\ and\ \bibinfo {author} {\bibfnamefont {H.}~\bibnamefont
  {Fehske}},\ }\bibfield  {title} {\bibinfo {title} {Route to {Chaos} in
  {Optomechanics}},\ }\href {https://doi.org/10.1103/PhysRevLett.114.013601}
  {\bibfield  {journal} {\bibinfo  {journal} {Physical Review Letters}\
  }\textbf {\bibinfo {volume} {114}},\ \bibinfo {pages} {013601} (\bibinfo
  {year} {2015})}\BibitemShut {NoStop}%
\bibitem [{\citenamefont {Carmon}\ \emph {et~al.}(2007)\citenamefont {Carmon},
  \citenamefont {Cross},\ and\ \citenamefont {Vahala}}]{carmon2007}%
  \BibitemOpen
  \bibfield  {author} {\bibinfo {author} {\bibfnamefont {T.}~\bibnamefont
  {Carmon}}, \bibinfo {author} {\bibfnamefont {M.~C.}\ \bibnamefont {Cross}},\
  and\ \bibinfo {author} {\bibfnamefont {K.~J.}\ \bibnamefont {Vahala}},\
  }\bibfield  {title} {\bibinfo {title} {Chaotic {Quivering} of
  {Micron}-{Scaled} {On}-{Chip} {Resonators} {Excited} by {Centrifugal}
  {Optical} {Pressure}},\ }\href
  {https://doi.org/10.1103/PhysRevLett.98.167203} {\bibfield  {journal}
  {\bibinfo  {journal} {Physical Review Letters}\ }\textbf {\bibinfo {volume}
  {98}},\ \bibinfo {pages} {167203} (\bibinfo {year} {2007})}\BibitemShut
  {NoStop}%
\bibitem [{\citenamefont {Zhang}\ and\ \citenamefont {Kong}(2014)}]{zhang2014}%
  \BibitemOpen
  \bibfield  {author} {\bibinfo {author} {\bibfnamefont {L.}~\bibnamefont
  {Zhang}}\ and\ \bibinfo {author} {\bibfnamefont {H.-Y.}\ \bibnamefont
  {Kong}},\ }\bibfield  {title} {\bibinfo {title} {Self-sustained oscillation
  and harmonic generation in optomechanical systems with quadratic couplings},\
  }\href {https://doi.org/10.1103/PhysRevA.89.023847} {\bibfield  {journal}
  {\bibinfo  {journal} {Physical Review A}\ }\textbf {\bibinfo {volume} {89}},\
  \bibinfo {pages} {023847} (\bibinfo {year} {2014})}\BibitemShut {NoStop}%
\bibitem [{\citenamefont {Zhang}\ \emph {et~al.}(2017)\citenamefont {Zhang},
  \citenamefont {Ji}, \citenamefont {Zhang},\ and\ \citenamefont
  {Zhang}}]{zhang2017}%
  \BibitemOpen
  \bibfield  {author} {\bibinfo {author} {\bibfnamefont {L.}~\bibnamefont
  {Zhang}}, \bibinfo {author} {\bibfnamefont {F.}~\bibnamefont {Ji}}, \bibinfo
  {author} {\bibfnamefont {X.}~\bibnamefont {Zhang}},\ and\ \bibinfo {author}
  {\bibfnamefont {W.}~\bibnamefont {Zhang}},\ }\bibfield  {title} {\bibinfo
  {title} {Photon–phonon parametric oscillation induced by quadratic coupling
  in an optomechanical resonator},\ }\href
  {https://doi.org/10.1088/1361-6455/aa74a0} {\bibfield  {journal} {\bibinfo
  {journal} {Journal of Physics B: Atomic, Molecular and Optical Physics}\
  }\textbf {\bibinfo {volume} {50}},\ \bibinfo {pages} {145501} (\bibinfo
  {year} {2017})}\BibitemShut {NoStop}%
\bibitem [{\citenamefont {Wurl}\ \emph {et~al.}(2016)\citenamefont {Wurl},
  \citenamefont {Alvermann},\ and\ \citenamefont {Fehske}}]{wurl2016}%
  \BibitemOpen
  \bibfield  {author} {\bibinfo {author} {\bibfnamefont {C.}~\bibnamefont
  {Wurl}}, \bibinfo {author} {\bibfnamefont {A.}~\bibnamefont {Alvermann}},\
  and\ \bibinfo {author} {\bibfnamefont {H.}~\bibnamefont {Fehske}},\
  }\bibfield  {title} {\bibinfo {title} {Symmetry-breaking oscillations in
  membrane optomechanics},\ }\href {https://doi.org/10.1103/PhysRevA.94.063860}
  {\bibfield  {journal} {\bibinfo  {journal} {Physical Review A}\ }\textbf
  {\bibinfo {volume} {94}},\ \bibinfo {pages} {063860} (\bibinfo {year}
  {2016})}\BibitemShut {NoStop}%
\bibitem [{\citenamefont {Mumford}\ \emph {et~al.}(2015)\citenamefont
  {Mumford}, \citenamefont {O'Dell},\ and\ \citenamefont
  {Larson}}]{mumford2015}%
  \BibitemOpen
  \bibfield  {author} {\bibinfo {author} {\bibfnamefont {J.}~\bibnamefont
  {Mumford}}, \bibinfo {author} {\bibfnamefont {D.~H.~J.}\ \bibnamefont
  {O'Dell}},\ and\ \bibinfo {author} {\bibfnamefont {J.}~\bibnamefont
  {Larson}},\ }\bibfield  {title} {\bibinfo {title} {Dicke-type phase
  transition in a multimode optomechanical system},\ }\href
  {https://doi.org/10.1002/andp.201400105} {\bibfield  {journal} {\bibinfo
  {journal} {Annalen der Physik}\ }\textbf {\bibinfo {volume} {527}},\ \bibinfo
  {pages} {115} (\bibinfo {year} {2015})}\BibitemShut {NoStop}%
\bibitem [{\citenamefont {Lü}\ \emph {et~al.}(2015)\citenamefont {Lü},
  \citenamefont {Jing}, \citenamefont {Ma},\ and\ \citenamefont {Wu}}]{lu2015}%
  \BibitemOpen
  \bibfield  {author} {\bibinfo {author} {\bibfnamefont {X.-Y.}\ \bibnamefont
  {Lü}}, \bibinfo {author} {\bibfnamefont {H.}~\bibnamefont {Jing}}, \bibinfo
  {author} {\bibfnamefont {J.-Y.}\ \bibnamefont {Ma}},\ and\ \bibinfo {author}
  {\bibfnamefont {Y.}~\bibnamefont {Wu}},\ }\bibfield  {title} {\bibinfo
  {title} {$\mathcal{P}\mathcal{T}$-{Symmetry}-{Breaking} {Chaos} in
  {Optomechanics}},\ }\href {https://doi.org/10.1103/PhysRevLett.114.253601}
  {\bibfield  {journal} {\bibinfo  {journal} {Physical Review Letters}\
  }\textbf {\bibinfo {volume} {114}},\ \bibinfo {pages} {253601} (\bibinfo
  {year} {2015})}\BibitemShut {NoStop}%
\bibitem [{\citenamefont {Roque}\ \emph {et~al.}(2020)\citenamefont {Roque},
  \citenamefont {Marquardt},\ and\ \citenamefont {Yevtushenko}}]{roque2020}%
  \BibitemOpen
  \bibfield  {author} {\bibinfo {author} {\bibfnamefont {T.~F.}\ \bibnamefont
  {Roque}}, \bibinfo {author} {\bibfnamefont {F.}~\bibnamefont {Marquardt}},\
  and\ \bibinfo {author} {\bibfnamefont {O.~M.}\ \bibnamefont {Yevtushenko}},\
  }\bibfield  {title} {\bibinfo {title} {Nonlinear dynamics of weakly
  dissipative optomechanical systems},\ }\href
  {https://doi.org/10.1088/1367-2630/ab6522} {\bibfield  {journal} {\bibinfo
  {journal} {New Journal of Physics}\ }\textbf {\bibinfo {volume} {22}},\
  \bibinfo {pages} {013049} (\bibinfo {year} {2020})}\BibitemShut {NoStop}%
\bibitem [{\citenamefont {Zhang}\ \emph {et~al.}(2020)\citenamefont {Zhang},
  \citenamefont {You},\ and\ \citenamefont {Lü}}]{zhang2020}%
  \BibitemOpen
  \bibfield  {author} {\bibinfo {author} {\bibfnamefont {D.-W.}\ \bibnamefont
  {Zhang}}, \bibinfo {author} {\bibfnamefont {C.}~\bibnamefont {You}},\ and\
  \bibinfo {author} {\bibfnamefont {X.-Y.}\ \bibnamefont {Lü}},\ }\bibfield
  {title} {\bibinfo {title} {Intermittent chaos in cavity optomechanics},\
  }\href {https://doi.org/10.1103/PhysRevA.101.053851} {\bibfield  {journal}
  {\bibinfo  {journal} {Physical Review A}\ }\textbf {\bibinfo {volume}
  {101}},\ \bibinfo {pages} {053851} (\bibinfo {year} {2020})}\BibitemShut
  {NoStop}%
\bibitem [{\citenamefont {Christou}\ \emph {et~al.}(2021)\citenamefont
  {Christou}, \citenamefont {Kovanis}, \citenamefont {Giannakopoulos},\ and\
  \citenamefont {Kominis}}]{christou2021}%
  \BibitemOpen
  \bibfield  {author} {\bibinfo {author} {\bibfnamefont {S.}~\bibnamefont
  {Christou}}, \bibinfo {author} {\bibfnamefont {V.}~\bibnamefont {Kovanis}},
  \bibinfo {author} {\bibfnamefont {A.~E.}\ \bibnamefont {Giannakopoulos}},\
  and\ \bibinfo {author} {\bibfnamefont {Y.}~\bibnamefont {Kominis}},\
  }\bibfield  {title} {\bibinfo {title} {Parametric control of self-sustained
  and self-modulated optomechanical oscillations},\ }\href
  {https://doi.org/10.1103/PhysRevA.103.053513} {\bibfield  {journal} {\bibinfo
   {journal} {Physical Review A}\ }\textbf {\bibinfo {volume} {103}},\ \bibinfo
  {pages} {053513} (\bibinfo {year} {2021})}\BibitemShut {NoStop}%
\bibitem [{\citenamefont {Ludwig}\ \emph {et~al.}(2008)\citenamefont {Ludwig},
  \citenamefont {Kubala},\ and\ \citenamefont {Marquardt}}]{ludwig2008}%
  \BibitemOpen
  \bibfield  {author} {\bibinfo {author} {\bibfnamefont {M.}~\bibnamefont
  {Ludwig}}, \bibinfo {author} {\bibfnamefont {B.}~\bibnamefont {Kubala}},\
  and\ \bibinfo {author} {\bibfnamefont {F.}~\bibnamefont {Marquardt}},\
  }\bibfield  {title} {\bibinfo {title} {The optomechanical instability in the
  quantum regime},\ }\href {https://doi.org/10.1088/1367-2630/10/9/095013}
  {\bibfield  {journal} {\bibinfo  {journal} {New Journal of Physics}\ }\textbf
  {\bibinfo {volume} {10}},\ \bibinfo {pages} {095013} (\bibinfo {year}
  {2008})}\BibitemShut {NoStop}%
\bibitem [{\citenamefont {Schulz}\ \emph {et~al.}(2016)\citenamefont {Schulz},
  \citenamefont {Alvermann}, \citenamefont {Bakemeier},\ and\ \citenamefont
  {Fehske}}]{schulz2016}%
  \BibitemOpen
  \bibfield  {author} {\bibinfo {author} {\bibfnamefont {C.}~\bibnamefont
  {Schulz}}, \bibinfo {author} {\bibfnamefont {A.}~\bibnamefont {Alvermann}},
  \bibinfo {author} {\bibfnamefont {L.}~\bibnamefont {Bakemeier}},\ and\
  \bibinfo {author} {\bibfnamefont {H.}~\bibnamefont {Fehske}},\ }\bibfield
  {title} {\bibinfo {title} {Optomechanical multistability in the quantum
  regime},\ }\href {https://doi.org/10.1209/0295-5075/113/64002} {\bibfield
  {journal} {\bibinfo  {journal} {Europhysics Letters}\ }\textbf {\bibinfo
  {volume} {113}},\ \bibinfo {pages} {64002} (\bibinfo {year}
  {2016})}\BibitemShut {NoStop}%
\bibitem [{\citenamefont {Monifi}\ \emph {et~al.}(2016)\citenamefont {Monifi},
  \citenamefont {Zhang}, \citenamefont {\"Ozdemir}, \citenamefont {Peng},
  \citenamefont {Liu}, \citenamefont {Bo}, \citenamefont {Nori},\ and\
  \citenamefont {Yang}}]{monifi2016}%
  \BibitemOpen
  \bibfield  {author} {\bibinfo {author} {\bibfnamefont {F.}~\bibnamefont
  {Monifi}}, \bibinfo {author} {\bibfnamefont {J.}~\bibnamefont {Zhang}},
  \bibinfo {author} {\bibfnamefont {c.~K.}\ \bibnamefont {\"Ozdemir}}, \bibinfo
  {author} {\bibfnamefont {B.}~\bibnamefont {Peng}}, \bibinfo {author}
  {\bibfnamefont {Y.-x.}\ \bibnamefont {Liu}}, \bibinfo {author} {\bibfnamefont
  {F.}~\bibnamefont {Bo}}, \bibinfo {author} {\bibfnamefont {F.}~\bibnamefont
  {Nori}},\ and\ \bibinfo {author} {\bibfnamefont {L.}~\bibnamefont {Yang}},\
  }\bibfield  {title} {\bibinfo {title} {Optomechanically induced stochastic
  resonance and chaos transfer between optical fields},\ }\href
  {https://doi.org/10.1038/nphoton.2016.73} {\bibfield  {journal} {\bibinfo
  {journal} {Nature Photonics}\ }\textbf {\bibinfo {volume} {10}},\ \bibinfo
  {pages} {399} (\bibinfo {year} {2016})}\BibitemShut {NoStop}%
\bibitem [{\citenamefont {Navarro-Urrios}\ \emph {et~al.}(2017)\citenamefont
  {Navarro-Urrios}, \citenamefont {Capuj}, \citenamefont {Colombano},
  \citenamefont {García}, \citenamefont {Sledzinska}, \citenamefont {Alzina},
  \citenamefont {Griol}, \citenamefont {Martínez},\ and\ \citenamefont
  {Sotomayor-Torres}}]{navarro-urrios2017}%
  \BibitemOpen
  \bibfield  {author} {\bibinfo {author} {\bibfnamefont {D.}~\bibnamefont
  {Navarro-Urrios}}, \bibinfo {author} {\bibfnamefont {N.~E.}\ \bibnamefont
  {Capuj}}, \bibinfo {author} {\bibfnamefont {M.~F.}\ \bibnamefont
  {Colombano}}, \bibinfo {author} {\bibfnamefont {P.~D.}\ \bibnamefont
  {García}}, \bibinfo {author} {\bibfnamefont {M.}~\bibnamefont {Sledzinska}},
  \bibinfo {author} {\bibfnamefont {F.}~\bibnamefont {Alzina}}, \bibinfo
  {author} {\bibfnamefont {A.}~\bibnamefont {Griol}}, \bibinfo {author}
  {\bibfnamefont {A.}~\bibnamefont {Martínez}},\ and\ \bibinfo {author}
  {\bibfnamefont {C.~M.}\ \bibnamefont {Sotomayor-Torres}},\ }\bibfield
  {title} {\bibinfo {title} {Nonlinear dynamics and chaos in an optomechanical
  beam},\ }\href {https://doi.org/10.1038/ncomms14965} {\bibfield  {journal}
  {\bibinfo  {journal} {Nature Communications}\ }\textbf {\bibinfo {volume}
  {8}},\ \bibinfo {pages} {14965} (\bibinfo {year} {2017})}\BibitemShut
  {NoStop}%
\bibitem [{\citenamefont {Wu}\ \emph {et~al.}(2017)\citenamefont {Wu},
  \citenamefont {Huang}, \citenamefont {Huang}, \citenamefont {Zhou},
  \citenamefont {Yang}, \citenamefont {Liu}, \citenamefont {Yu}, \citenamefont
  {Lo}, \citenamefont {Kwong}, \citenamefont {Duan},\ and\ \citenamefont
  {Wei~Wong}}]{wu2017}%
  \BibitemOpen
  \bibfield  {author} {\bibinfo {author} {\bibfnamefont {J.}~\bibnamefont
  {Wu}}, \bibinfo {author} {\bibfnamefont {S.-W.}\ \bibnamefont {Huang}},
  \bibinfo {author} {\bibfnamefont {Y.}~\bibnamefont {Huang}}, \bibinfo
  {author} {\bibfnamefont {H.}~\bibnamefont {Zhou}}, \bibinfo {author}
  {\bibfnamefont {J.}~\bibnamefont {Yang}}, \bibinfo {author} {\bibfnamefont
  {J.-M.}\ \bibnamefont {Liu}}, \bibinfo {author} {\bibfnamefont
  {M.}~\bibnamefont {Yu}}, \bibinfo {author} {\bibfnamefont {G.}~\bibnamefont
  {Lo}}, \bibinfo {author} {\bibfnamefont {D.-L.}\ \bibnamefont {Kwong}},
  \bibinfo {author} {\bibfnamefont {S.}~\bibnamefont {Duan}},\ and\ \bibinfo
  {author} {\bibfnamefont {C.}~\bibnamefont {Wei~Wong}},\ }\bibfield  {title}
  {\bibinfo {title} {Mesoscopic chaos mediated by {Drude} electron-hole plasma
  in silicon optomechanical oscillators},\ }\href
  {https://doi.org/10.1038/ncomms15570} {\bibfield  {journal} {\bibinfo
  {journal} {Nature Communications}\ }\textbf {\bibinfo {volume} {8}},\
  \bibinfo {pages} {15570} (\bibinfo {year} {2017})}\BibitemShut {NoStop}%
\bibitem [{\citenamefont {Madiot}\ \emph {et~al.}(2021)\citenamefont {Madiot},
  \citenamefont {Correia}, \citenamefont {Barbay},\ and\ \citenamefont
  {Braive}}]{madiot2021}%
  \BibitemOpen
  \bibfield  {author} {\bibinfo {author} {\bibfnamefont {G.}~\bibnamefont
  {Madiot}}, \bibinfo {author} {\bibfnamefont {F.}~\bibnamefont {Correia}},
  \bibinfo {author} {\bibfnamefont {S.}~\bibnamefont {Barbay}},\ and\ \bibinfo
  {author} {\bibfnamefont {R.}~\bibnamefont {Braive}},\ }\bibfield  {title}
  {\bibinfo {title} {Bichromatic synchronized chaos in driven coupled
  electro-optomechanical nanoresonators},\ }\href
  {https://doi.org/10.1103/PhysRevA.104.023525} {\bibfield  {journal} {\bibinfo
   {journal} {Physical Review A}\ }\textbf {\bibinfo {volume} {104}},\ \bibinfo
  {pages} {023525} (\bibinfo {year} {2021})}\BibitemShut {NoStop}%
\bibitem [{\citenamefont {Rossler}(1983)}]{rossler1983}%
  \BibitemOpen
  \bibfield  {author} {\bibinfo {author} {\bibfnamefont {O.~E.}\ \bibnamefont
  {Rossler}},\ }\bibfield  {title} {\bibinfo {title} {The {Chaotic}
  {Hierarchy}},\ }\href@noop {} {\bibfield  {journal} {\bibinfo  {journal}
  {Zeitschrift f\"ur Naturforschung A}\ }\textbf {\bibinfo {volume} {38}},\
  \bibinfo {pages} {788} (\bibinfo {year} {1983})}\BibitemShut {NoStop}%
\bibitem [{\citenamefont {Rossler}(1979)}]{rossler1979}%
  \BibitemOpen
  \bibfield  {author} {\bibinfo {author} {\bibfnamefont {O.~E.}\ \bibnamefont
  {Rossler}},\ }\bibfield  {title} {\bibinfo {title} {An equation for
  hyperchaos},\ }\href {https://doi.org/10.1016/0375-9601(79)90150-6}
  {\bibfield  {journal} {\bibinfo  {journal} {Physics Letters A}\ }\textbf
  {\bibinfo {volume} {71}},\ \bibinfo {pages} {155} (\bibinfo {year}
  {1979})}\BibitemShut {NoStop}%
\bibitem [{\citenamefont {Kapitaniak}\ and\ \citenamefont
  {Steeb}(1991)}]{kapitaniak1991}%
  \BibitemOpen
  \bibfield  {author} {\bibinfo {author} {\bibfnamefont {T.}~\bibnamefont
  {Kapitaniak}}\ and\ \bibinfo {author} {\bibfnamefont {W.~H.}\ \bibnamefont
  {Steeb}},\ }\bibfield  {title} {\bibinfo {title} {Transition to hyperchaos in
  coupled generalized van der {Pol} equations},\ }\href
  {https://doi.org/10.1016/0375-9601(91)90624-H} {\bibfield  {journal}
  {\bibinfo  {journal} {Physics Letters A}\ }\textbf {\bibinfo {volume}
  {152}},\ \bibinfo {pages} {33} (\bibinfo {year} {1991})}\BibitemShut
  {NoStop}%
\bibitem [{\citenamefont {Harrison}\ and\ \citenamefont
  {Lai}(1999)}]{harrison1999}%
  \BibitemOpen
  \bibfield  {author} {\bibinfo {author} {\bibfnamefont {M.~A.}\ \bibnamefont
  {Harrison}}\ and\ \bibinfo {author} {\bibfnamefont {Y.-C.}\ \bibnamefont
  {Lai}},\ }\bibfield  {title} {\bibinfo {title} {Route to high-dimensional
  chaos},\ }\href {https://doi.org/10.1103/PhysRevE.59.R3799} {\bibfield
  {journal} {\bibinfo  {journal} {Physical Review E}\ }\textbf {\bibinfo
  {volume} {59}},\ \bibinfo {pages} {R3799} (\bibinfo {year}
  {1999})}\BibitemShut {NoStop}%
\bibitem [{\citenamefont {Kapitaniak}\ \emph {et~al.}(2000)\citenamefont
  {Kapitaniak}, \citenamefont {Maistrenko},\ and\ \citenamefont
  {Popovych}}]{kapitaniak2000}%
  \BibitemOpen
  \bibfield  {author} {\bibinfo {author} {\bibfnamefont {T.}~\bibnamefont
  {Kapitaniak}}, \bibinfo {author} {\bibfnamefont {Y.}~\bibnamefont
  {Maistrenko}},\ and\ \bibinfo {author} {\bibfnamefont {S.}~\bibnamefont
  {Popovych}},\ }\bibfield  {title} {\bibinfo {title} {Chaos-hyperchaos
  transition},\ }\href {https://doi.org/10.1103/PhysRevE.62.1972} {\bibfield
  {journal} {\bibinfo  {journal} {Physical Review E}\ }\textbf {\bibinfo
  {volume} {62}},\ \bibinfo {pages} {1972} (\bibinfo {year}
  {2000})}\BibitemShut {NoStop}%
\bibitem [{\citenamefont {Pavlov}\ \emph {et~al.}(2015)\citenamefont {Pavlov},
  \citenamefont {Pavlova}, \citenamefont {Mohammad},\ and\ \citenamefont
  {Kurths}}]{pavlov2015}%
  \BibitemOpen
  \bibfield  {author} {\bibinfo {author} {\bibfnamefont {A.~N.}\ \bibnamefont
  {Pavlov}}, \bibinfo {author} {\bibfnamefont {O.~N.}\ \bibnamefont {Pavlova}},
  \bibinfo {author} {\bibfnamefont {Y.~K.}\ \bibnamefont {Mohammad}},\ and\
  \bibinfo {author} {\bibfnamefont {J.}~\bibnamefont {Kurths}},\ }\bibfield
  {title} {\bibinfo {title} {Characterization of the chaos-hyperchaos
  transition based on return times},\ }\href
  {https://doi.org/10.1103/PhysRevE.91.022921} {\bibfield  {journal} {\bibinfo
  {journal} {Physical Review E}\ }\textbf {\bibinfo {volume} {91}},\ \bibinfo
  {pages} {022921} (\bibinfo {year} {2015})}\BibitemShut {NoStop}%
\bibitem [{\citenamefont {Du}\ \emph {et~al.}(2018)\citenamefont {Du},
  \citenamefont {Qu},\ and\ \citenamefont {Lai}}]{du2018}%
  \BibitemOpen
  \bibfield  {author} {\bibinfo {author} {\bibfnamefont {R.-H.}\ \bibnamefont
  {Du}}, \bibinfo {author} {\bibfnamefont {S.-X.}\ \bibnamefont {Qu}},\ and\
  \bibinfo {author} {\bibfnamefont {Y.-C.}\ \bibnamefont {Lai}},\ }\bibfield
  {title} {\bibinfo {title} {Transition to high-dimensional chaos in nonsmooth
  dynamical systems},\ }\href {https://doi.org/10.1103/PhysRevE.98.052212}
  {\bibfield  {journal} {\bibinfo  {journal} {Physical Review E}\ }\textbf
  {\bibinfo {volume} {98}},\ \bibinfo {pages} {052212} (\bibinfo {year}
  {2018})}\BibitemShut {NoStop}%
\bibitem [{\citenamefont {Leo~Kingston}\ \emph {et~al.}(2023)\citenamefont
  {Leo~Kingston}, \citenamefont {Balcerzak}, \citenamefont {Dana},\ and\
  \citenamefont {Kapitaniak}}]{leo_kingston2023}%
  \BibitemOpen
  \bibfield  {author} {\bibinfo {author} {\bibfnamefont {S.}~\bibnamefont
  {Leo~Kingston}}, \bibinfo {author} {\bibfnamefont {M.}~\bibnamefont
  {Balcerzak}}, \bibinfo {author} {\bibfnamefont {S.~K.}\ \bibnamefont
  {Dana}},\ and\ \bibinfo {author} {\bibfnamefont {T.}~\bibnamefont
  {Kapitaniak}},\ }\bibfield  {title} {\bibinfo {title} {Transition to
  hyperchaos and rare large-intensity pulses in {Zeeman} laser},\ }\href
  {https://doi.org/10.1063/5.0135228} {\bibfield  {journal} {\bibinfo
  {journal} {Chaos: An Interdisciplinary Journal of Nonlinear Science}\
  }\textbf {\bibinfo {volume} {33}},\ \bibinfo {pages} {023128} (\bibinfo
  {year} {2023})}\BibitemShut {NoStop}%
\bibitem [{\citenamefont {Matsumoto}\ \emph {et~al.}(1986)\citenamefont
  {Matsumoto}, \citenamefont {Chua},\ and\ \citenamefont
  {Kobayashi}}]{matsumoto1986}%
  \BibitemOpen
  \bibfield  {author} {\bibinfo {author} {\bibfnamefont {T.}~\bibnamefont
  {Matsumoto}}, \bibinfo {author} {\bibfnamefont {L.}~\bibnamefont {Chua}},\
  and\ \bibinfo {author} {\bibfnamefont {K.}~\bibnamefont {Kobayashi}},\
  }\bibfield  {title} {\bibinfo {title} {Hyper chaos: {Laboratory} experiment
  and numerical confirmation},\ }\href
  {https://doi.org/10.1109/TCS.1986.1085862} {\bibfield  {journal} {\bibinfo
  {journal} {IEEE Transactions on Circuits and Systems}\ }\textbf {\bibinfo
  {volume} {33}},\ \bibinfo {pages} {1143} (\bibinfo {year}
  {1986})}\BibitemShut {NoStop}%
\bibitem [{\citenamefont {Stoop}\ and\ \citenamefont
  {Meier}(1988)}]{stoop1988}%
  \BibitemOpen
  \bibfield  {author} {\bibinfo {author} {\bibfnamefont {R.}~\bibnamefont
  {Stoop}}\ and\ \bibinfo {author} {\bibfnamefont {P.~F.}\ \bibnamefont
  {Meier}},\ }\bibfield  {title} {\bibinfo {title} {Evaluation of {Lyapunov}
  exponents and scaling functions from time series},\ }\href
  {https://doi.org/10.1364/JOSAB.5.001037} {\bibfield  {journal} {\bibinfo
  {journal} {JOSA B}\ }\textbf {\bibinfo {volume} {5}},\ \bibinfo {pages}
  {1037} (\bibinfo {year} {1988})}\BibitemShut {NoStop}%
\bibitem [{\citenamefont {Stoop}\ \emph {et~al.}(1989)\citenamefont {Stoop},
  \citenamefont {Peinke}, \citenamefont {Parisi}, \citenamefont {R\"ohricht},\
  and\ \citenamefont {Huebener}}]{stoop1989}%
  \BibitemOpen
  \bibfield  {author} {\bibinfo {author} {\bibfnamefont {R.}~\bibnamefont
  {Stoop}}, \bibinfo {author} {\bibfnamefont {J.}~\bibnamefont {Peinke}},
  \bibinfo {author} {\bibfnamefont {J.}~\bibnamefont {Parisi}}, \bibinfo
  {author} {\bibfnamefont {B.}~\bibnamefont {R\"ohricht}},\ and\ \bibinfo
  {author} {\bibfnamefont {R.~P.}\ \bibnamefont {Huebener}},\ }\bibfield
  {title} {\bibinfo {title} {A p-{Ge} semiconductor experiment showing chaos
  and hyperchaos},\ }\href {https://doi.org/10.1016/0167-2789(89)90078-X}
  {\bibfield  {journal} {\bibinfo  {journal} {Physica D: Nonlinear Phenomena}\
  }\textbf {\bibinfo {volume} {35}},\ \bibinfo {pages} {425} (\bibinfo {year}
  {1989})}\BibitemShut {NoStop}%
\bibitem [{\citenamefont {Eiswirth}\ \emph {et~al.}(1992)\citenamefont
  {Eiswirth}, \citenamefont {Kruel}, \citenamefont {Ertl},\ and\ \citenamefont
  {Schneider}}]{eiswirth1992}%
  \BibitemOpen
  \bibfield  {author} {\bibinfo {author} {\bibfnamefont {M.}~\bibnamefont
  {Eiswirth}}, \bibinfo {author} {\bibfnamefont {T.~M.}\ \bibnamefont {Kruel}},
  \bibinfo {author} {\bibfnamefont {G.}~\bibnamefont {Ertl}},\ and\ \bibinfo
  {author} {\bibfnamefont {F.~W.}\ \bibnamefont {Schneider}},\ }\bibfield
  {title} {\bibinfo {title} {Hyperchaos in a chemical reaction},\ }\href
  {https://doi.org/10.1016/0009-2614(92)85672-W} {\bibfield  {journal}
  {\bibinfo  {journal} {Chemical Physics Letters}\ }\textbf {\bibinfo {volume}
  {193}},\ \bibinfo {pages} {305} (\bibinfo {year} {1992})}\BibitemShut
  {NoStop}%
\bibitem [{\citenamefont {Fischer}\ \emph {et~al.}(1994)\citenamefont
  {Fischer}, \citenamefont {Hess}, \citenamefont {Els\"a\ss{}er},\ and\
  \citenamefont {G\"obel}}]{fischer1994}%
  \BibitemOpen
  \bibfield  {author} {\bibinfo {author} {\bibfnamefont {I.}~\bibnamefont
  {Fischer}}, \bibinfo {author} {\bibfnamefont {O.}~\bibnamefont {Hess}},
  \bibinfo {author} {\bibfnamefont {W.}~\bibnamefont {Els\"a\ss{}er}},\ and\
  \bibinfo {author} {\bibfnamefont {E.}~\bibnamefont {G\"obel}},\ }\bibfield
  {title} {\bibinfo {title} {High-{Dimensional} {Chaotic} {Dynamics} of an
  {External} {Cavity} {Semiconductor} {Laser}},\ }\href
  {https://doi.org/10.1103/PhysRevLett.73.2188} {\bibfield  {journal} {\bibinfo
   {journal} {Physical Review Letters}\ }\textbf {\bibinfo {volume} {73}},\
  \bibinfo {pages} {2188} (\bibinfo {year} {1994})}\BibitemShut {NoStop}%
\bibitem [{\citenamefont {Deng}\ \emph {et~al.}(2022)\citenamefont {Deng},
  \citenamefont {Fan}, \citenamefont {Zhao}, \citenamefont {Wang},
  \citenamefont {Zhao}, \citenamefont {Wu}, \citenamefont {Grillot},\ and\
  \citenamefont {Wang}}]{deng2022}%
  \BibitemOpen
  \bibfield  {author} {\bibinfo {author} {\bibfnamefont {Y.}~\bibnamefont
  {Deng}}, \bibinfo {author} {\bibfnamefont {Z.-F.}\ \bibnamefont {Fan}},
  \bibinfo {author} {\bibfnamefont {B.-B.}\ \bibnamefont {Zhao}}, \bibinfo
  {author} {\bibfnamefont {X.-G.}\ \bibnamefont {Wang}}, \bibinfo {author}
  {\bibfnamefont {S.}~\bibnamefont {Zhao}}, \bibinfo {author} {\bibfnamefont
  {J.}~\bibnamefont {Wu}}, \bibinfo {author} {\bibfnamefont {F.}~\bibnamefont
  {Grillot}},\ and\ \bibinfo {author} {\bibfnamefont {C.}~\bibnamefont
  {Wang}},\ }\bibfield  {title} {\bibinfo {title} {Mid-infrared hyperchaos of
  interband cascade lasers},\ }\href
  {https://doi.org/10.1038/s41377-021-00697-1} {\bibfield  {journal} {\bibinfo
  {journal} {Light: Science \& Applications}\ }\textbf {\bibinfo {volume}
  {11}},\ \bibinfo {pages} {7} (\bibinfo {year} {2022})}\BibitemShut {NoStop}%
\bibitem [{\citenamefont {Bir}\ \emph {et~al.}(2020)\citenamefont {Bir},
  \citenamefont {Grishin}, \citenamefont {Moskalenko}, \citenamefont {Pavlov},
  \citenamefont {Zhuravlev},\ and\ \citenamefont {Ruiz}}]{bir2020}%
  \BibitemOpen
  \bibfield  {author} {\bibinfo {author} {\bibfnamefont {A.~S.}\ \bibnamefont
  {Bir}}, \bibinfo {author} {\bibfnamefont {S.~V.}\ \bibnamefont {Grishin}},
  \bibinfo {author} {\bibfnamefont {O.~I.}\ \bibnamefont {Moskalenko}},
  \bibinfo {author} {\bibfnamefont {A.~N.}\ \bibnamefont {Pavlov}}, \bibinfo
  {author} {\bibfnamefont {M.~O.}\ \bibnamefont {Zhuravlev}},\ and\ \bibinfo
  {author} {\bibfnamefont {D.~O.}\ \bibnamefont {Ruiz}},\ }\bibfield  {title}
  {\bibinfo {title} {Experimental {Observation} of {Ultrashort} {Hyperchaotic}
  {Dark} {Multisoliton} {Complexes} in a {Magnonic} {Active} {Ring}
  {Resonator}},\ }\href {https://doi.org/10.1103/PhysRevLett.125.083903}
  {\bibfield  {journal} {\bibinfo  {journal} {Physical Review Letters}\
  }\textbf {\bibinfo {volume} {125}},\ \bibinfo {pages} {083903} (\bibinfo
  {year} {2020})}\BibitemShut {NoStop}%
\bibitem [{\citenamefont {Abarbanel}\ \emph {et~al.}(1993)\citenamefont
  {Abarbanel}, \citenamefont {Brown}, \citenamefont {Sidorowich},\ and\
  \citenamefont {Tsimring}}]{abarbanel1993}%
  \BibitemOpen
  \bibfield  {author} {\bibinfo {author} {\bibfnamefont {H.~D.~I.}\
  \bibnamefont {Abarbanel}}, \bibinfo {author} {\bibfnamefont {R.}~\bibnamefont
  {Brown}}, \bibinfo {author} {\bibfnamefont {J.~J.}\ \bibnamefont
  {Sidorowich}},\ and\ \bibinfo {author} {\bibfnamefont {L.~S.}\ \bibnamefont
  {Tsimring}},\ }\bibfield  {title} {\bibinfo {title} {The analysis of observed
  chaotic data in physical systems},\ }\href
  {https://doi.org/10.1103/RevModPhys.65.1331} {\bibfield  {journal} {\bibinfo
  {journal} {Reviews of Modern Physics}\ }\textbf {\bibinfo {volume} {65}},\
  \bibinfo {pages} {1331} (\bibinfo {year} {1993})}\BibitemShut {NoStop}%
\bibitem [{\citenamefont {Bonatto}(2018)}]{bonatto2018}%
  \BibitemOpen
  \bibfield  {author} {\bibinfo {author} {\bibfnamefont {C.}~\bibnamefont
  {Bonatto}},\ }\bibfield  {title} {\bibinfo {title} {Hyperchaotic {Dynamics}
  for {Light} {Polarization} in a {Laser} {Diode}},\ }\href
  {https://doi.org/10.1103/PhysRevLett.120.163902} {\bibfield  {journal}
  {\bibinfo  {journal} {Physical Review Letters}\ }\textbf {\bibinfo {volume}
  {120}},\ \bibinfo {pages} {163902} (\bibinfo {year} {2018})}\BibitemShut
  {NoStop}%
\bibitem [{\citenamefont {Mompó}\ \emph {et~al.}(2021)\citenamefont {Mompó},
  \citenamefont {Carretero},\ and\ \citenamefont {Bonilla}}]{mompo2021}%
  \BibitemOpen
  \bibfield  {author} {\bibinfo {author} {\bibfnamefont {E.}~\bibnamefont
  {Mompó}}, \bibinfo {author} {\bibfnamefont {M.}~\bibnamefont {Carretero}},\
  and\ \bibinfo {author} {\bibfnamefont {L.}~\bibnamefont {Bonilla}},\
  }\bibfield  {title} {\bibinfo {title} {Designing {Hyperchaos} and
  {Intermittency} in {Semiconductor} {Superlattices}},\ }\href
  {https://doi.org/10.1103/PhysRevLett.127.096601} {\bibfield  {journal}
  {\bibinfo  {journal} {Physical Review Letters}\ }\textbf {\bibinfo {volume}
  {127}},\ \bibinfo {pages} {096601} (\bibinfo {year} {2021})}\BibitemShut
  {NoStop}%
\bibitem [{\citenamefont {Macek}\ and\ \citenamefont
  {Strumik}(2014)}]{macek2014}%
  \BibitemOpen
  \bibfield  {author} {\bibinfo {author} {\bibfnamefont {W.~M.}\ \bibnamefont
  {Macek}}\ and\ \bibinfo {author} {\bibfnamefont {M.}~\bibnamefont
  {Strumik}},\ }\bibfield  {title} {\bibinfo {title} {Hyperchaotic
  {Intermittent} {Convection} in a {Magnetized} {Viscous} {Fluid}},\ }\href
  {https://doi.org/10.1103/PhysRevLett.112.074502} {\bibfield  {journal}
  {\bibinfo  {journal} {Physical Review Letters}\ }\textbf {\bibinfo {volume}
  {112}},\ \bibinfo {pages} {074502} (\bibinfo {year} {2014})}\BibitemShut
  {NoStop}%
\bibitem [{\citenamefont {Andreev}\ \emph {et~al.}(2021)\citenamefont
  {Andreev}, \citenamefont {Balanov}, \citenamefont {Fromhold}, \citenamefont
  {Greenaway}, \citenamefont {Hramov}, \citenamefont {Li}, \citenamefont
  {Makarov},\ and\ \citenamefont {Zagoskin}}]{andreev2021}%
  \BibitemOpen
  \bibfield  {author} {\bibinfo {author} {\bibfnamefont {A.~V.}\ \bibnamefont
  {Andreev}}, \bibinfo {author} {\bibfnamefont {A.~G.}\ \bibnamefont
  {Balanov}}, \bibinfo {author} {\bibfnamefont {T.~M.}\ \bibnamefont
  {Fromhold}}, \bibinfo {author} {\bibfnamefont {M.~T.}\ \bibnamefont
  {Greenaway}}, \bibinfo {author} {\bibfnamefont {A.~E.}\ \bibnamefont
  {Hramov}}, \bibinfo {author} {\bibfnamefont {W.}~\bibnamefont {Li}}, \bibinfo
  {author} {\bibfnamefont {V.~V.}\ \bibnamefont {Makarov}},\ and\ \bibinfo
  {author} {\bibfnamefont {A.~M.}\ \bibnamefont {Zagoskin}},\ }\bibfield
  {title} {\bibinfo {title} {Emergence and control of complex behaviors in
  driven systems of interacting qubits with dissipation},\ }\href
  {https://doi.org/10.1038/s41534-020-00339-1} {\bibfield  {journal} {\bibinfo
  {journal} {npj Quantum Information}\ }\textbf {\bibinfo {volume} {7}},\
  \bibinfo {pages} {1} (\bibinfo {year} {2021})}\BibitemShut {NoStop}%
\bibitem [{\citenamefont {Sankey}\ \emph {et~al.}(2010)\citenamefont {Sankey},
  \citenamefont {Yang}, \citenamefont {Zwickl}, \citenamefont {Jayich},\ and\
  \citenamefont {Harris}}]{sankey2010}%
  \BibitemOpen
  \bibfield  {author} {\bibinfo {author} {\bibfnamefont {J.~C.}\ \bibnamefont
  {Sankey}}, \bibinfo {author} {\bibfnamefont {C.}~\bibnamefont {Yang}},
  \bibinfo {author} {\bibfnamefont {B.~M.}\ \bibnamefont {Zwickl}}, \bibinfo
  {author} {\bibfnamefont {A.~M.}\ \bibnamefont {Jayich}},\ and\ \bibinfo
  {author} {\bibfnamefont {J.~G.~E.}\ \bibnamefont {Harris}},\ }\bibfield
  {title} {\bibinfo {title} {Strong and tunable nonlinear optomechanical
  coupling in a low-loss system},\ }\href {https://doi.org/10.1038/nphys1707}
  {\bibfield  {journal} {\bibinfo  {journal} {Nature Physics}\ }\textbf
  {\bibinfo {volume} {6}},\ \bibinfo {pages} {707} (\bibinfo {year}
  {2010})}\BibitemShut {NoStop}%
\bibitem [{\citenamefont {Para\"{\i}so}\ \emph {et~al.}(2015)\citenamefont
  {Para\"{\i}so}, \citenamefont {Kalaee}, \citenamefont {Zang}, \citenamefont
  {Pfeifer}, \citenamefont {Marquardt},\ and\ \citenamefont
  {Painter}}]{paraiso2015}%
  \BibitemOpen
  \bibfield  {author} {\bibinfo {author} {\bibfnamefont {T.~K.}\ \bibnamefont
  {Para\"{\i}so}}, \bibinfo {author} {\bibfnamefont {M.}~\bibnamefont
  {Kalaee}}, \bibinfo {author} {\bibfnamefont {L.}~\bibnamefont {Zang}},
  \bibinfo {author} {\bibfnamefont {H.}~\bibnamefont {Pfeifer}}, \bibinfo
  {author} {\bibfnamefont {F.}~\bibnamefont {Marquardt}},\ and\ \bibinfo
  {author} {\bibfnamefont {O.}~\bibnamefont {Painter}},\ }\bibfield  {title}
  {\bibinfo {title} {Position-{Squared} {Coupling} in a {Tunable} {Photonic}
  {Crystal} {Optomechanical} {Cavity}},\ }\href
  {https://doi.org/10.1103/PhysRevX.5.041024} {\bibfield  {journal} {\bibinfo
  {journal} {Physical Review X}\ }\textbf {\bibinfo {volume} {5}},\ \bibinfo
  {pages} {041024} (\bibinfo {year} {2015})}\BibitemShut {NoStop}%
\bibitem [{\citenamefont {Kaviani}\ \emph {et~al.}(2015)\citenamefont
  {Kaviani}, \citenamefont {Healey}, \citenamefont {Wu}, \citenamefont
  {Ghobadi}, \citenamefont {Hryciw},\ and\ \citenamefont
  {Barclay}}]{kaviani2015}%
  \BibitemOpen
  \bibfield  {author} {\bibinfo {author} {\bibfnamefont {H.}~\bibnamefont
  {Kaviani}}, \bibinfo {author} {\bibfnamefont {C.}~\bibnamefont {Healey}},
  \bibinfo {author} {\bibfnamefont {M.}~\bibnamefont {Wu}}, \bibinfo {author}
  {\bibfnamefont {R.}~\bibnamefont {Ghobadi}}, \bibinfo {author} {\bibfnamefont
  {A.}~\bibnamefont {Hryciw}},\ and\ \bibinfo {author} {\bibfnamefont {P.~E.}\
  \bibnamefont {Barclay}},\ }\bibfield  {title} {\bibinfo {title} {Nonlinear
  optomechanical paddle nanocavities},\ }\href
  {https://doi.org/10.1364/OPTICA.2.000271} {\bibfield  {journal} {\bibinfo
  {journal} {Optica}\ }\textbf {\bibinfo {volume} {2}},\ \bibinfo {pages} {271}
  (\bibinfo {year} {2015})}\BibitemShut {NoStop}%
\bibitem [{\citenamefont {Burgwal}\ \emph {et~al.}(2020)\citenamefont
  {Burgwal}, \citenamefont {Pino},\ and\ \citenamefont
  {Verhagen}}]{burgwal2020}%
  \BibitemOpen
  \bibfield  {author} {\bibinfo {author} {\bibfnamefont {R.}~\bibnamefont
  {Burgwal}}, \bibinfo {author} {\bibfnamefont {J.~d.}\ \bibnamefont {Pino}},\
  and\ \bibinfo {author} {\bibfnamefont {E.}~\bibnamefont {Verhagen}},\
  }\bibfield  {title} {\bibinfo {title} {Comparing nonlinear optomechanical
  coupling in membrane-in-the-middle and single-cavity systems},\ }\href
  {https://doi.org/10.1088/1367-2630/abc1c8} {\bibfield  {journal} {\bibinfo
  {journal} {New Journal of Physics}\ }\textbf {\bibinfo {volume} {22}},\
  \bibinfo {pages} {113006} (\bibinfo {year} {2020})}\BibitemShut {NoStop}%
\bibitem [{\citenamefont {Bullier}\ \emph {et~al.}(2021)\citenamefont
  {Bullier}, \citenamefont {Pontin},\ and\ \citenamefont
  {Barker}}]{bullier2021}%
  \BibitemOpen
  \bibfield  {author} {\bibinfo {author} {\bibfnamefont {N.~P.}\ \bibnamefont
  {Bullier}}, \bibinfo {author} {\bibfnamefont {A.}~\bibnamefont {Pontin}},\
  and\ \bibinfo {author} {\bibfnamefont {P.~F.}\ \bibnamefont {Barker}},\
  }\bibfield  {title} {\bibinfo {title} {Quadratic optomechanical cooling of a
  cavity-levitated nanosphere},\ }\href
  {https://doi.org/10.1103/PhysRevResearch.3.L032022} {\bibfield  {journal}
  {\bibinfo  {journal} {Physical Review Research}\ }\textbf {\bibinfo {volume}
  {3}},\ \bibinfo {pages} {L032022} (\bibinfo {year} {2021})}\BibitemShut
  {NoStop}%
\bibitem [{\citenamefont {Burgwal}\ and\ \citenamefont
  {Verhagen}(2023)}]{burgwal2023}%
  \BibitemOpen
  \bibfield  {author} {\bibinfo {author} {\bibfnamefont {R.}~\bibnamefont
  {Burgwal}}\ and\ \bibinfo {author} {\bibfnamefont {E.}~\bibnamefont
  {Verhagen}},\ }\bibfield  {title} {\bibinfo {title} {Enhanced nonlinear
  optomechanics in a coupled-mode photonic crystal device},\ }\href
  {https://doi.org/10.1038/s41467-023-37138-z} {\bibfield  {journal} {\bibinfo
  {journal} {Nature Communications}\ }\textbf {\bibinfo {volume} {14}},\
  \bibinfo {pages} {1526} (\bibinfo {year} {2023})}\BibitemShut {NoStop}%
\bibitem [{\citenamefont {Thompson}\ \emph {et~al.}(2008)\citenamefont
  {Thompson}, \citenamefont {Zwickl}, \citenamefont {Jayich}, \citenamefont
  {Marquardt}, \citenamefont {Girvin},\ and\ \citenamefont
  {Harris}}]{thompson2008}%
  \BibitemOpen
  \bibfield  {author} {\bibinfo {author} {\bibfnamefont {J.~D.}\ \bibnamefont
  {Thompson}}, \bibinfo {author} {\bibfnamefont {B.~M.}\ \bibnamefont
  {Zwickl}}, \bibinfo {author} {\bibfnamefont {A.~M.}\ \bibnamefont {Jayich}},
  \bibinfo {author} {\bibfnamefont {F.}~\bibnamefont {Marquardt}}, \bibinfo
  {author} {\bibfnamefont {S.~M.}\ \bibnamefont {Girvin}},\ and\ \bibinfo
  {author} {\bibfnamefont {J.~G.~E.}\ \bibnamefont {Harris}},\ }\bibfield
  {title} {\bibinfo {title} {Strong dispersive coupling of a high-finesse
  cavity to a micromechanical membrane},\ }\href
  {https://doi.org/10.1038/nature06715} {\bibfield  {journal} {\bibinfo
  {journal} {Nature}\ }\textbf {\bibinfo {volume} {452}},\ \bibinfo {pages}
  {72} (\bibinfo {year} {2008})}\BibitemShut {NoStop}%
\bibitem [{\citenamefont {Jayich}\ \emph {et~al.}(2008)\citenamefont {Jayich},
  \citenamefont {Sankey}, \citenamefont {Zwickl}, \citenamefont {Yang},
  \citenamefont {Thompson}, \citenamefont {Girvin}, \citenamefont {Clerk},
  \citenamefont {Marquardt},\ and\ \citenamefont {Harris}}]{jayich2008}%
  \BibitemOpen
  \bibfield  {author} {\bibinfo {author} {\bibfnamefont {A.~M.}\ \bibnamefont
  {Jayich}}, \bibinfo {author} {\bibfnamefont {J.~C.}\ \bibnamefont {Sankey}},
  \bibinfo {author} {\bibfnamefont {B.~M.}\ \bibnamefont {Zwickl}}, \bibinfo
  {author} {\bibfnamefont {C.}~\bibnamefont {Yang}}, \bibinfo {author}
  {\bibfnamefont {J.~D.}\ \bibnamefont {Thompson}}, \bibinfo {author}
  {\bibfnamefont {S.~M.}\ \bibnamefont {Girvin}}, \bibinfo {author}
  {\bibfnamefont {A.~A.}\ \bibnamefont {Clerk}}, \bibinfo {author}
  {\bibfnamefont {F.}~\bibnamefont {Marquardt}},\ and\ \bibinfo {author}
  {\bibfnamefont {J.~G.~E.}\ \bibnamefont {Harris}},\ }\bibfield  {title}
  {\bibinfo {title} {Dispersive optomechanics: a membrane inside a cavity},\
  }\href {https://doi.org/10.1088/1367-2630/10/9/095008} {\bibfield  {journal}
  {\bibinfo  {journal} {New Journal of Physics}\ }\textbf {\bibinfo {volume}
  {10}},\ \bibinfo {pages} {095008} (\bibinfo {year} {2008})}\BibitemShut
  {NoStop}%
\bibitem [{\citenamefont {Wilson}\ \emph {et~al.}(2009)\citenamefont {Wilson},
  \citenamefont {Regal}, \citenamefont {Papp},\ and\ \citenamefont
  {Kimble}}]{wilson2009}%
  \BibitemOpen
  \bibfield  {author} {\bibinfo {author} {\bibfnamefont {D.~J.}\ \bibnamefont
  {Wilson}}, \bibinfo {author} {\bibfnamefont {C.~A.}\ \bibnamefont {Regal}},
  \bibinfo {author} {\bibfnamefont {S.~B.}\ \bibnamefont {Papp}},\ and\
  \bibinfo {author} {\bibfnamefont {H.~J.}\ \bibnamefont {Kimble}},\ }\bibfield
   {title} {\bibinfo {title} {Cavity {Optomechanics} with {Stoichiometric}
  {SiN} {Films}},\ }\href {https://doi.org/10.1103/PhysRevLett.103.207204}
  {\bibfield  {journal} {\bibinfo  {journal} {Physical Review Letters}\
  }\textbf {\bibinfo {volume} {103}},\ \bibinfo {pages} {207204} (\bibinfo
  {year} {2009})}\BibitemShut {NoStop}%
\bibitem [{\citenamefont {Cheung}\ and\ \citenamefont
  {Law}(2011)}]{cheung2011}%
  \BibitemOpen
  \bibfield  {author} {\bibinfo {author} {\bibfnamefont {H.~K.}\ \bibnamefont
  {Cheung}}\ and\ \bibinfo {author} {\bibfnamefont {C.~K.}\ \bibnamefont
  {Law}},\ }\bibfield  {title} {\bibinfo {title} {Nonadiabatic optomechanical
  {Hamiltonian} of a moving dielectric membrane in a cavity},\ }\href
  {https://doi.org/10.1103/PhysRevA.84.023812} {\bibfield  {journal} {\bibinfo
  {journal} {Physical Review A}\ }\textbf {\bibinfo {volume} {84}},\ \bibinfo
  {pages} {023812} (\bibinfo {year} {2011})}\BibitemShut {NoStop}%
\bibitem [{Note1()}]{Note1}%
  \BibitemOpen
  \bibinfo {note} {If $G<0$ the minus sign appears outside the square root in
  $\protect \tilde x,\protect \tilde p$. The equations of motion are invariant
  under $x\rightarrow -x$, $p\rightarrow -p$.}\BibitemShut {Stop}%
\bibitem [{\citenamefont {Buchmann}\ \emph {et~al.}(2012)\citenamefont
  {Buchmann}, \citenamefont {Zhang}, \citenamefont {Chiruvelli},\ and\
  \citenamefont {Meystre}}]{buchmann2012}%
  \BibitemOpen
  \bibfield  {author} {\bibinfo {author} {\bibfnamefont {L.~F.}\ \bibnamefont
  {Buchmann}}, \bibinfo {author} {\bibfnamefont {L.}~\bibnamefont {Zhang}},
  \bibinfo {author} {\bibfnamefont {A.}~\bibnamefont {Chiruvelli}},\ and\
  \bibinfo {author} {\bibfnamefont {P.}~\bibnamefont {Meystre}},\ }\bibfield
  {title} {\bibinfo {title} {Macroscopic {Tunneling} of a {Membrane} in an
  {Optomechanical} {Double}-{Well} {Potential}},\ }\href
  {https://doi.org/10.1103/PhysRevLett.108.210403} {\bibfield  {journal}
  {\bibinfo  {journal} {Physical Review Letters}\ }\textbf {\bibinfo {volume}
  {108}},\ \bibinfo {pages} {210403} (\bibinfo {year} {2012})}\BibitemShut
  {NoStop}%
\bibitem [{\citenamefont {Seok}\ \emph {et~al.}(2013)\citenamefont {Seok},
  \citenamefont {Buchmann}, \citenamefont {Wright},\ and\ \citenamefont
  {Meystre}}]{seok2013}%
  \BibitemOpen
  \bibfield  {author} {\bibinfo {author} {\bibfnamefont {H.}~\bibnamefont
  {Seok}}, \bibinfo {author} {\bibfnamefont {L.~F.}\ \bibnamefont {Buchmann}},
  \bibinfo {author} {\bibfnamefont {E.~M.}\ \bibnamefont {Wright}},\ and\
  \bibinfo {author} {\bibfnamefont {P.}~\bibnamefont {Meystre}},\ }\bibfield
  {title} {\bibinfo {title} {Multimode strong-coupling quantum optomechanics},\
  }\href {https://doi.org/10.1103/PhysRevA.88.063850} {\bibfield  {journal}
  {\bibinfo  {journal} {Physical Review A}\ }\textbf {\bibinfo {volume} {88}},\
  \bibinfo {pages} {063850} (\bibinfo {year} {2013})}\BibitemShut {NoStop}%
\bibitem [{\citenamefont {Strogatz}(2015)}]{strogatz2015}%
  \BibitemOpen
  \bibfield  {author} {\bibinfo {author} {\bibfnamefont {S.~H.}\ \bibnamefont
  {Strogatz}},\ }\href@noop {} {\emph {\bibinfo {title} {Nonlinear dynamics and
  chaos: with applications to physics, biology, chemistry, and engineering}}},\
  \bibinfo {edition} {second edition}\ ed.\ (\bibinfo  {publisher} {Westview
  Press, a member of the Perseus Books Group},\ \bibinfo {address} {Boulder,
  CO},\ \bibinfo {year} {2015})\BibitemShut {NoStop}%
\bibitem [{\citenamefont {Wolf}\ \emph {et~al.}(1985)\citenamefont {Wolf},
  \citenamefont {Swift}, \citenamefont {Swinney},\ and\ \citenamefont
  {Vastano}}]{wolf1985}%
  \BibitemOpen
  \bibfield  {author} {\bibinfo {author} {\bibfnamefont {A.}~\bibnamefont
  {Wolf}}, \bibinfo {author} {\bibfnamefont {J.~B.}\ \bibnamefont {Swift}},
  \bibinfo {author} {\bibfnamefont {H.~L.}\ \bibnamefont {Swinney}},\ and\
  \bibinfo {author} {\bibfnamefont {J.~A.}\ \bibnamefont {Vastano}},\
  }\bibfield  {title} {\bibinfo {title} {Determining {Lyapunov} exponents from
  a time series},\ }\href {https://doi.org/10.1016/0167-2789(85)90011-9}
  {\bibfield  {journal} {\bibinfo  {journal} {Physica D: Nonlinear Phenomena}\
  }\textbf {\bibinfo {volume} {16}},\ \bibinfo {pages} {285} (\bibinfo {year}
  {1985})}\BibitemShut {NoStop}%
\bibitem [{\citenamefont {Bryant}\ \emph {et~al.}(1990)\citenamefont {Bryant},
  \citenamefont {Brown},\ and\ \citenamefont {Abarbanel}}]{bryant1990}%
  \BibitemOpen
  \bibfield  {author} {\bibinfo {author} {\bibfnamefont {P.}~\bibnamefont
  {Bryant}}, \bibinfo {author} {\bibfnamefont {R.}~\bibnamefont {Brown}},\ and\
  \bibinfo {author} {\bibfnamefont {H.~D.~I.}\ \bibnamefont {Abarbanel}},\
  }\bibfield  {title} {\bibinfo {title} {Lyapunov exponents from observed time
  series},\ }\href {https://doi.org/10.1103/PhysRevLett.65.1523} {\bibfield
  {journal} {\bibinfo  {journal} {Physical Review Letters}\ }\textbf {\bibinfo
  {volume} {65}},\ \bibinfo {pages} {1523} (\bibinfo {year}
  {1990})}\BibitemShut {NoStop}%
\bibitem [{\citenamefont {Grassberger}\ and\ \citenamefont
  {Procaccia}(1983)}]{grassberger1983}%
  \BibitemOpen
  \bibfield  {author} {\bibinfo {author} {\bibfnamefont {P.}~\bibnamefont
  {Grassberger}}\ and\ \bibinfo {author} {\bibfnamefont {I.}~\bibnamefont
  {Procaccia}},\ }\bibfield  {title} {\bibinfo {title} {Measuring the
  strangeness of strange attractors},\ }\href
  {https://doi.org/10.1016/0167-2789(83)90298-1} {\bibfield  {journal}
  {\bibinfo  {journal} {Physica D: Nonlinear Phenomena}\ }\textbf {\bibinfo
  {volume} {9}},\ \bibinfo {pages} {189} (\bibinfo {year} {1983})}\BibitemShut
  {NoStop}%
\bibitem [{\citenamefont {Sauer}\ \emph {et~al.}(1997)\citenamefont {Sauer},
  \citenamefont {Grebogi},\ and\ \citenamefont {Yorke}}]{sauer1997}%
  \BibitemOpen
  \bibfield  {author} {\bibinfo {author} {\bibfnamefont {T.}~\bibnamefont
  {Sauer}}, \bibinfo {author} {\bibfnamefont {C.}~\bibnamefont {Grebogi}},\
  and\ \bibinfo {author} {\bibfnamefont {J.~A.}\ \bibnamefont {Yorke}},\
  }\bibfield  {title} {\bibinfo {title} {How {Long} {Do} {Numerical} {Chaotic}
  {Solutions} {Remain} {Valid}?},\ }\href
  {https://doi.org/10.1103/PhysRevLett.79.59} {\bibfield  {journal} {\bibinfo
  {journal} {Physical Review Letters}\ }\textbf {\bibinfo {volume} {79}},\
  \bibinfo {pages} {59} (\bibinfo {year} {1997})}\BibitemShut {NoStop}%
\bibitem [{\citenamefont {Grebogi}\ \emph {et~al.}(1990)\citenamefont
  {Grebogi}, \citenamefont {Hammel}, \citenamefont {Yorke},\ and\ \citenamefont
  {Sauer}}]{grebogi1990}%
  \BibitemOpen
  \bibfield  {author} {\bibinfo {author} {\bibfnamefont {C.}~\bibnamefont
  {Grebogi}}, \bibinfo {author} {\bibfnamefont {S.~M.}\ \bibnamefont {Hammel}},
  \bibinfo {author} {\bibfnamefont {J.~A.}\ \bibnamefont {Yorke}},\ and\
  \bibinfo {author} {\bibfnamefont {T.}~\bibnamefont {Sauer}},\ }\bibfield
  {title} {\bibinfo {title} {Shadowing of physical trajectories in chaotic
  dynamics: {Containment} and refinement},\ }\href
  {https://doi.org/10.1103/PhysRevLett.65.1527} {\bibfield  {journal} {\bibinfo
   {journal} {Physical Review Letters}\ }\textbf {\bibinfo {volume} {65}},\
  \bibinfo {pages} {1527} (\bibinfo {year} {1990})}\BibitemShut {NoStop}%
\bibitem [{\citenamefont {McCullen}\ and\ \citenamefont
  {Moresco}(2011)}]{mccullen2011}%
  \BibitemOpen
  \bibfield  {author} {\bibinfo {author} {\bibfnamefont {N.~J.}\ \bibnamefont
  {McCullen}}\ and\ \bibinfo {author} {\bibfnamefont {P.}~\bibnamefont
  {Moresco}},\ }\bibfield  {title} {\bibinfo {title} {Route to hyperchaos in a
  system of coupled oscillators with multistability},\ }\href
  {https://doi.org/10.1103/PhysRevE.83.046212} {\bibfield  {journal} {\bibinfo
  {journal} {Physical Review E}\ }\textbf {\bibinfo {volume} {83}},\ \bibinfo
  {pages} {046212} (\bibinfo {year} {2011})}\BibitemShut {NoStop}%
\bibitem [{\citenamefont {Davidchack}\ and\ \citenamefont
  {Lai}(2000)}]{davidchack2000}%
  \BibitemOpen
  \bibfield  {author} {\bibinfo {author} {\bibfnamefont {R.}~\bibnamefont
  {Davidchack}}\ and\ \bibinfo {author} {\bibfnamefont {Y.-C.}\ \bibnamefont
  {Lai}},\ }\bibfield  {title} {\bibinfo {title} {Characterization of
  transition to chaos with multiple positive {Lyapunov} exponents by unstable
  periodic orbits},\ }\href {https://doi.org/10.1016/S0375-9601(00)00335-2}
  {\bibfield  {journal} {\bibinfo  {journal} {Physics Letters A}\ }\textbf
  {\bibinfo {volume} {270}},\ \bibinfo {pages} {308} (\bibinfo {year}
  {2000})}\BibitemShut {NoStop}%
\bibitem [{\citenamefont {Auerbach}\ \emph {et~al.}(1987)\citenamefont
  {Auerbach}, \citenamefont {Cvitanović}, \citenamefont {Eckmann},
  \citenamefont {Gunaratne},\ and\ \citenamefont {Procaccia}}]{auerbach1987}%
  \BibitemOpen
  \bibfield  {author} {\bibinfo {author} {\bibfnamefont {D.}~\bibnamefont
  {Auerbach}}, \bibinfo {author} {\bibfnamefont {P.}~\bibnamefont
  {Cvitanović}}, \bibinfo {author} {\bibfnamefont {J.-P.}\ \bibnamefont
  {Eckmann}}, \bibinfo {author} {\bibfnamefont {G.}~\bibnamefont {Gunaratne}},\
  and\ \bibinfo {author} {\bibfnamefont {I.}~\bibnamefont {Procaccia}},\
  }\bibfield  {title} {\bibinfo {title} {Exploring chaotic motion through
  periodic orbits},\ }\href {https://doi.org/10.1103/PhysRevLett.58.2387}
  {\bibfield  {journal} {\bibinfo  {journal} {Physical Review Letters}\
  }\textbf {\bibinfo {volume} {58}},\ \bibinfo {pages} {2387} (\bibinfo {year}
  {1987})}\BibitemShut {NoStop}%
\bibitem [{\citenamefont {Dawson}(1996)}]{dawson1996}%
  \BibitemOpen
  \bibfield  {author} {\bibinfo {author} {\bibfnamefont {S.~P.}\ \bibnamefont
  {Dawson}},\ }\bibfield  {title} {\bibinfo {title} {Strange {Nonattracting}
  {Chaotic} {Sets}, {Crises}, and {Fluctuating} {Lyapunov} {Exponents}},\
  }\href {https://doi.org/10.1103/PhysRevLett.76.4348} {\bibfield  {journal}
  {\bibinfo  {journal} {Physical Review Letters}\ }\textbf {\bibinfo {volume}
  {76}},\ \bibinfo {pages} {4348} (\bibinfo {year} {1996})}\BibitemShut
  {NoStop}%
\bibitem [{\citenamefont {Dawson}\ \emph {et~al.}(1994)\citenamefont {Dawson},
  \citenamefont {Grebogi}, \citenamefont {Sauer},\ and\ \citenamefont
  {Yorke}}]{dawson1994}%
  \BibitemOpen
  \bibfield  {author} {\bibinfo {author} {\bibfnamefont {S.}~\bibnamefont
  {Dawson}}, \bibinfo {author} {\bibfnamefont {C.}~\bibnamefont {Grebogi}},
  \bibinfo {author} {\bibfnamefont {T.}~\bibnamefont {Sauer}},\ and\ \bibinfo
  {author} {\bibfnamefont {J.~A.}\ \bibnamefont {Yorke}},\ }\bibfield  {title}
  {\bibinfo {title} {Obstructions to {Shadowing} {When} a {Lyapunov} {Exponent}
  {Fluctuates} about {Zero}},\ }\href
  {https://doi.org/10.1103/PhysRevLett.73.1927} {\bibfield  {journal} {\bibinfo
   {journal} {Physical Review Letters}\ }\textbf {\bibinfo {volume} {73}},\
  \bibinfo {pages} {1927} (\bibinfo {year} {1994})}\BibitemShut {NoStop}%
\bibitem [{\citenamefont {Lai}\ \emph {et~al.}(1999)\citenamefont {Lai},
  \citenamefont {Lerner}, \citenamefont {Williams},\ and\ \citenamefont
  {Grebogi}}]{lai1999}%
  \BibitemOpen
  \bibfield  {author} {\bibinfo {author} {\bibfnamefont {Y.-C.}\ \bibnamefont
  {Lai}}, \bibinfo {author} {\bibfnamefont {D.}~\bibnamefont {Lerner}},
  \bibinfo {author} {\bibfnamefont {K.}~\bibnamefont {Williams}},\ and\
  \bibinfo {author} {\bibfnamefont {C.}~\bibnamefont {Grebogi}},\ }\bibfield
  {title} {\bibinfo {title} {Unstable dimension variability in coupled chaotic
  systems},\ }\href {https://doi.org/10.1103/PhysRevE.60.5445} {\bibfield
  {journal} {\bibinfo  {journal} {Physical Review E}\ }\textbf {\bibinfo
  {volume} {60}},\ \bibinfo {pages} {5445} (\bibinfo {year}
  {1999})}\BibitemShut {NoStop}%
\bibitem [{\citenamefont {Barreto}\ and\ \citenamefont
  {So}(2000)}]{barreto2000}%
  \BibitemOpen
  \bibfield  {author} {\bibinfo {author} {\bibfnamefont {E.}~\bibnamefont
  {Barreto}}\ and\ \bibinfo {author} {\bibfnamefont {P.}~\bibnamefont {So}},\
  }\bibfield  {title} {\bibinfo {title} {Mechanisms for the {Development} of
  {Unstable} {Dimension} {Variability} and the {Breakdown} of {Shadowing} in
  {Coupled} {Chaotic} {Systems}},\ }\href
  {https://doi.org/10.1103/PhysRevLett.85.2490} {\bibfield  {journal} {\bibinfo
   {journal} {Physical Review Letters}\ }\textbf {\bibinfo {volume} {85}},\
  \bibinfo {pages} {2490} (\bibinfo {year} {2000})}\BibitemShut {NoStop}%
\bibitem [{\citenamefont {Kuptsov}(2013)}]{kuptsov2013}%
  \BibitemOpen
  \bibfield  {author} {\bibinfo {author} {\bibfnamefont {P.~V.}\ \bibnamefont
  {Kuptsov}},\ }\bibfield  {title} {\bibinfo {title} {Violation of
  hyperbolicity via unstable dimension variability in a chain with local
  hyperbolic chaotic attractors},\ }\href
  {https://doi.org/10.1088/1751-8113/46/25/254016} {\bibfield  {journal}
  {\bibinfo  {journal} {Journal of Physics A: Mathematical and Theoretical}\
  }\textbf {\bibinfo {volume} {46}},\ \bibinfo {pages} {254016} (\bibinfo
  {year} {2013})}\BibitemShut {NoStop}%
\bibitem [{\citenamefont {Elste}\ \emph {et~al.}(2009)\citenamefont {Elste},
  \citenamefont {Girvin},\ and\ \citenamefont {Clerk}}]{elste2009}%
  \BibitemOpen
  \bibfield  {author} {\bibinfo {author} {\bibfnamefont {F.}~\bibnamefont
  {Elste}}, \bibinfo {author} {\bibfnamefont {S.~M.}\ \bibnamefont {Girvin}},\
  and\ \bibinfo {author} {\bibfnamefont {A.~A.}\ \bibnamefont {Clerk}},\
  }\bibfield  {title} {\bibinfo {title} {Quantum {Noise} {Interference} and
  {Backaction} {Cooling} in {Cavity} {Nanomechanics}},\ }\href
  {https://doi.org/10.1103/PhysRevLett.102.207209} {\bibfield  {journal}
  {\bibinfo  {journal} {Physical Review Letters}\ }\textbf {\bibinfo {volume}
  {102}},\ \bibinfo {pages} {207209} (\bibinfo {year} {2009})}\BibitemShut
  {NoStop}%
\bibitem [{\citenamefont {Yanay}\ \emph {et~al.}(2016)\citenamefont {Yanay},
  \citenamefont {Sankey},\ and\ \citenamefont {Clerk}}]{yanay2016}%
  \BibitemOpen
  \bibfield  {author} {\bibinfo {author} {\bibfnamefont {Y.}~\bibnamefont
  {Yanay}}, \bibinfo {author} {\bibfnamefont {J.~C.}\ \bibnamefont {Sankey}},\
  and\ \bibinfo {author} {\bibfnamefont {A.~A.}\ \bibnamefont {Clerk}},\
  }\bibfield  {title} {\bibinfo {title} {Quantum backaction and noise
  interference in asymmetric two-cavity optomechanical systems},\ }\href
  {https://doi.org/10.1103/PhysRevA.93.063809} {\bibfield  {journal} {\bibinfo
  {journal} {Physical Review A}\ }\textbf {\bibinfo {volume} {93}},\ \bibinfo
  {pages} {063809} (\bibinfo {year} {2016})}\BibitemShut {NoStop}%
\bibitem [{\citenamefont {Reinhardt}\ \emph {et~al.}(2016)\citenamefont
  {Reinhardt}, \citenamefont {Müller}, \citenamefont {Bourassa},\ and\
  \citenamefont {Sankey}}]{reinhardt2016}%
  \BibitemOpen
  \bibfield  {author} {\bibinfo {author} {\bibfnamefont {C.}~\bibnamefont
  {Reinhardt}}, \bibinfo {author} {\bibfnamefont {T.}~\bibnamefont {Müller}},
  \bibinfo {author} {\bibfnamefont {A.}~\bibnamefont {Bourassa}},\ and\
  \bibinfo {author} {\bibfnamefont {J.~C.}\ \bibnamefont {Sankey}},\ }\bibfield
   {title} {\bibinfo {title} {Ultralow-{Noise} {SiN} {Trampoline} {Resonators}
  for {Sensing} and {Optomechanics}},\ }\href
  {https://doi.org/10.1103/PhysRevX.6.021001} {\bibfield  {journal} {\bibinfo
  {journal} {Physical Review X}\ }\textbf {\bibinfo {volume} {6}},\ \bibinfo
  {pages} {021001} (\bibinfo {year} {2016})}\BibitemShut {NoStop}%
\bibitem [{\citenamefont {Fonseca}\ \emph {et~al.}(2016)\citenamefont
  {Fonseca}, \citenamefont {Aranas}, \citenamefont {Millen}, \citenamefont
  {Monteiro},\ and\ \citenamefont {Barker}}]{fonseca2016}%
  \BibitemOpen
  \bibfield  {author} {\bibinfo {author} {\bibfnamefont {P.}~\bibnamefont
  {Fonseca}}, \bibinfo {author} {\bibfnamefont {E.}~\bibnamefont {Aranas}},
  \bibinfo {author} {\bibfnamefont {J.}~\bibnamefont {Millen}}, \bibinfo
  {author} {\bibfnamefont {T.}~\bibnamefont {Monteiro}},\ and\ \bibinfo
  {author} {\bibfnamefont {P.}~\bibnamefont {Barker}},\ }\bibfield  {title}
  {\bibinfo {title} {Nonlinear {Dynamics} and {Strong} {Cavity} {Cooling} of
  {Levitated} {Nanoparticles}},\ }\href
  {https://doi.org/10.1103/PhysRevLett.117.173602} {\bibfield  {journal}
  {\bibinfo  {journal} {Physical Review Letters}\ }\textbf {\bibinfo {volume}
  {117}},\ \bibinfo {pages} {173602} (\bibinfo {year} {2016})}\BibitemShut
  {NoStop}%
\bibitem [{\citenamefont {Delić}\ \emph {et~al.}(2020)\citenamefont {Delić},
  \citenamefont {Reisenbauer}, \citenamefont {Dare}, \citenamefont {Grass},
  \citenamefont {Vuletić}, \citenamefont {Kiesel},\ and\ \citenamefont
  {Aspelmeyer}}]{delic2020}%
  \BibitemOpen
  \bibfield  {author} {\bibinfo {author} {\bibfnamefont {U.}~\bibnamefont
  {Delić}}, \bibinfo {author} {\bibfnamefont {M.}~\bibnamefont {Reisenbauer}},
  \bibinfo {author} {\bibfnamefont {K.}~\bibnamefont {Dare}}, \bibinfo {author}
  {\bibfnamefont {D.}~\bibnamefont {Grass}}, \bibinfo {author} {\bibfnamefont
  {V.}~\bibnamefont {Vuletić}}, \bibinfo {author} {\bibfnamefont
  {N.}~\bibnamefont {Kiesel}},\ and\ \bibinfo {author} {\bibfnamefont
  {M.}~\bibnamefont {Aspelmeyer}},\ }\bibfield  {title} {\bibinfo {title}
  {Cooling of a levitated nanoparticle to the motional quantum ground state},\
  }\href {https://doi.org/10.1126/science.aba3993} {\bibfield  {journal}
  {\bibinfo  {journal} {Science}\ }\textbf {\bibinfo {volume} {367}},\ \bibinfo
  {pages} {892} (\bibinfo {year} {2020})}\BibitemShut {NoStop}%
\bibitem [{\citenamefont {Leo~Kingston}\ \emph {et~al.}(2022)\citenamefont
  {Leo~Kingston}, \citenamefont {Kapitaniak},\ and\ \citenamefont
  {Dana}}]{leo_kingston2022}%
  \BibitemOpen
  \bibfield  {author} {\bibinfo {author} {\bibfnamefont {S.}~\bibnamefont
  {Leo~Kingston}}, \bibinfo {author} {\bibfnamefont {T.}~\bibnamefont
  {Kapitaniak}},\ and\ \bibinfo {author} {\bibfnamefont {S.~K.}\ \bibnamefont
  {Dana}},\ }\bibfield  {title} {\bibinfo {title} {Transition to hyperchaos:
  {Sudden} expansion of attractor and intermittent large-amplitude events in
  dynamical systems},\ }\href {https://doi.org/10.1063/5.0108401} {\bibfield
  {journal} {\bibinfo  {journal} {Chaos: An Interdisciplinary Journal of
  Nonlinear Science}\ }\textbf {\bibinfo {volume} {32}},\ \bibinfo {pages}
  {081106} (\bibinfo {year} {2022})}\BibitemShut {NoStop}%
\bibitem [{\citenamefont {Garashchuk}\ \emph {et~al.}(2019)\citenamefont
  {Garashchuk}, \citenamefont {Sinelshchikov}, \citenamefont {Kazakov},\ and\
  \citenamefont {Kudryashov}}]{garashchuk2019}%
  \BibitemOpen
  \bibfield  {author} {\bibinfo {author} {\bibfnamefont {I.~R.}\ \bibnamefont
  {Garashchuk}}, \bibinfo {author} {\bibfnamefont {D.~I.}\ \bibnamefont
  {Sinelshchikov}}, \bibinfo {author} {\bibfnamefont {A.~O.}\ \bibnamefont
  {Kazakov}},\ and\ \bibinfo {author} {\bibfnamefont {N.~A.}\ \bibnamefont
  {Kudryashov}},\ }\bibfield  {title} {\bibinfo {title} {Hyperchaos and
  multistability in the model of two interacting microbubble contrast agents},\
  }\href {https://doi.org/10.1063/1.5098329} {\bibfield  {journal} {\bibinfo
  {journal} {Chaos: An Interdisciplinary Journal of Nonlinear Science}\
  }\textbf {\bibinfo {volume} {29}},\ \bibinfo {pages} {063131} (\bibinfo
  {year} {2019})}\BibitemShut {NoStop}%
\bibitem [{\citenamefont {Sajjadi}\ \emph {et~al.}(2020)\citenamefont
  {Sajjadi}, \citenamefont {Baleanu}, \citenamefont {Jajarmi},\ and\
  \citenamefont {Pirouz}}]{sajjadi2020}%
  \BibitemOpen
  \bibfield  {author} {\bibinfo {author} {\bibfnamefont {S.~S.}\ \bibnamefont
  {Sajjadi}}, \bibinfo {author} {\bibfnamefont {D.}~\bibnamefont {Baleanu}},
  \bibinfo {author} {\bibfnamefont {A.}~\bibnamefont {Jajarmi}},\ and\ \bibinfo
  {author} {\bibfnamefont {H.~M.}\ \bibnamefont {Pirouz}},\ }\bibfield  {title}
  {\bibinfo {title} {A new adaptive synchronization and hyperchaos control of a
  biological snap oscillator},\ }\href
  {https://doi.org/10.1016/j.chaos.2020.109919} {\bibfield  {journal} {\bibinfo
   {journal} {Chaos, Solitons \& Fractals}\ }\textbf {\bibinfo {volume}
  {138}},\ \bibinfo {pages} {109919} (\bibinfo {year} {2020})}\BibitemShut
  {NoStop}%
\bibitem [{\citenamefont {Lin}\ \emph {et~al.}(2020)\citenamefont {Lin},
  \citenamefont {Wang},\ and\ \citenamefont {Tan}}]{lin2020}%
  \BibitemOpen
  \bibfield  {author} {\bibinfo {author} {\bibfnamefont {H.}~\bibnamefont
  {Lin}}, \bibinfo {author} {\bibfnamefont {C.}~\bibnamefont {Wang}},\ and\
  \bibinfo {author} {\bibfnamefont {Y.}~\bibnamefont {Tan}},\ }\bibfield
  {title} {\bibinfo {title} {Hidden extreme multistability with hyperchaos and
  transient chaos in a {Hopfield} neural network affected by electromagnetic
  radiation},\ }\href {https://doi.org/10.1007/s11071-019-05408-5} {\bibfield
  {journal} {\bibinfo  {journal} {Nonlinear Dynamics}\ }\textbf {\bibinfo
  {volume} {99}},\ \bibinfo {pages} {2369} (\bibinfo {year}
  {2020})}\BibitemShut {NoStop}%
\bibitem [{\citenamefont {Datseris}\ and\ \citenamefont
  {Parlitz}(2022)}]{datseris2022}%
  \BibitemOpen
  \bibfield  {author} {\bibinfo {author} {\bibfnamefont {G.}~\bibnamefont
  {Datseris}}\ and\ \bibinfo {author} {\bibfnamefont {U.}~\bibnamefont
  {Parlitz}},\ }\href
  {https://link.springer.com/book/10.1007/978-3-030-91032-7} {\emph {\bibinfo
  {title} {Nonlinear dynamics: a concise introduction interlaced with code}}},\
  Undergraduate lecture notes in physics\ (\bibinfo  {publisher} {Springer
  Cham},\ \bibinfo {year} {2022})\BibitemShut {NoStop}%
\bibitem [{\citenamefont {Benettin}\ \emph
  {et~al.}(1980{\natexlab{a}})\citenamefont {Benettin}, \citenamefont
  {Galgani}, \citenamefont {Giorgilli},\ and\ \citenamefont
  {Strelcyn}}]{benettin1980}%
  \BibitemOpen
  \bibfield  {author} {\bibinfo {author} {\bibfnamefont {G.}~\bibnamefont
  {Benettin}}, \bibinfo {author} {\bibfnamefont {L.}~\bibnamefont {Galgani}},
  \bibinfo {author} {\bibfnamefont {A.}~\bibnamefont {Giorgilli}},\ and\
  \bibinfo {author} {\bibfnamefont {J.-M.}\ \bibnamefont {Strelcyn}},\
  }\bibfield  {title} {\bibinfo {title} {Lyapunov {Characteristic} {Exponents}
  for smooth dynamical systems and for hamiltonian systems; {A} method for
  computing all of them. {Part} 2: {Numerical} application},\ }\href
  {https://doi.org/10.1007/BF02128237} {\bibfield  {journal} {\bibinfo
  {journal} {Meccanica}\ }\textbf {\bibinfo {volume} {15}},\ \bibinfo {pages}
  {21} (\bibinfo {year} {1980}{\natexlab{a}})}\BibitemShut {NoStop}%
\bibitem [{\citenamefont {Benettin}\ \emph
  {et~al.}(1980{\natexlab{b}})\citenamefont {Benettin}, \citenamefont
  {Galgani}, \citenamefont {Giorgilli},\ and\ \citenamefont
  {Strelcyn}}]{benettin1980b}%
  \BibitemOpen
  \bibfield  {author} {\bibinfo {author} {\bibfnamefont {G.}~\bibnamefont
  {Benettin}}, \bibinfo {author} {\bibfnamefont {L.}~\bibnamefont {Galgani}},
  \bibinfo {author} {\bibfnamefont {A.}~\bibnamefont {Giorgilli}},\ and\
  \bibinfo {author} {\bibfnamefont {J.-M.}\ \bibnamefont {Strelcyn}},\
  }\bibfield  {title} {\bibinfo {title} {Lyapunov {Characteristic} {Exponents}
  for smooth dynamical systems and for hamiltonian systems; a method for
  computing all of them. {Part} 1: {Theory}},\ }\href
  {https://doi.org/10.1007/BF02128236} {\bibfield  {journal} {\bibinfo
  {journal} {Meccanica}\ }\textbf {\bibinfo {volume} {15}},\ \bibinfo {pages}
  {9} (\bibinfo {year} {1980}{\natexlab{b}})}\BibitemShut {NoStop}%
\bibitem [{\citenamefont {Datseris}\ \emph {et~al.}(2023)\citenamefont
  {Datseris}, \citenamefont {Kottlarz}, \citenamefont {Braun},\ and\
  \citenamefont {Parlitz}}]{datseris2023}%
  \BibitemOpen
  \bibfield  {author} {\bibinfo {author} {\bibfnamefont {G.}~\bibnamefont
  {Datseris}}, \bibinfo {author} {\bibfnamefont {I.}~\bibnamefont {Kottlarz}},
  \bibinfo {author} {\bibfnamefont {A.~P.}\ \bibnamefont {Braun}},\ and\
  \bibinfo {author} {\bibfnamefont {U.}~\bibnamefont {Parlitz}},\ }\bibfield
  {title} {\bibinfo {title} {Estimating fractal dimensions: {A} comparative
  review and open source implementations},\ }\href
  {https://doi.org/10.1063/5.0160394} {\bibfield  {journal} {\bibinfo
  {journal} {Chaos: An Interdisciplinary Journal of Nonlinear Science}\
  }\textbf {\bibinfo {volume} {33}},\ \bibinfo {pages} {102101} (\bibinfo
  {year} {2023})}\BibitemShut {NoStop}%
\bibitem [{\citenamefont {Judd}(1992)}]{judd1992}%
  \BibitemOpen
  \bibfield  {author} {\bibinfo {author} {\bibfnamefont {K.}~\bibnamefont
  {Judd}},\ }\bibfield  {title} {\bibinfo {title} {An improved estimator of
  dimension and some comments on providing confidence intervals},\ }\href
  {https://doi.org/10.1016/0167-2789(92)90025-I} {\bibfield  {journal}
  {\bibinfo  {journal} {Physica D: Nonlinear Phenomena}\ }\textbf {\bibinfo
  {volume} {56}},\ \bibinfo {pages} {216} (\bibinfo {year} {1992})}\BibitemShut
  {NoStop}%
\bibitem [{\citenamefont {Deshmukh}\ \emph {et~al.}(2021)\citenamefont
  {Deshmukh}, \citenamefont {Bradley}, \citenamefont {Garland},\ and\
  \citenamefont {Meiss}}]{deshmukh2021}%
  \BibitemOpen
  \bibfield  {author} {\bibinfo {author} {\bibfnamefont {V.}~\bibnamefont
  {Deshmukh}}, \bibinfo {author} {\bibfnamefont {E.}~\bibnamefont {Bradley}},
  \bibinfo {author} {\bibfnamefont {J.}~\bibnamefont {Garland}},\ and\ \bibinfo
  {author} {\bibfnamefont {J.~D.}\ \bibnamefont {Meiss}},\ }\bibfield  {title}
  {\bibinfo {title} {Toward automated extraction and characterization of
  scaling regions in dynamical systems},\ }\href
  {https://doi.org/10.1063/5.0069365} {\bibfield  {journal} {\bibinfo
  {journal} {Chaos: An Interdisciplinary Journal of Nonlinear Science}\
  }\textbf {\bibinfo {volume} {31}},\ \bibinfo {pages} {123102} (\bibinfo
  {year} {2021})}\BibitemShut {NoStop}%
\end{thebibliography}%

\end{document}